\documentclass[sigconf]{acmart}
\pdfoutput=1

\renewcommand\footnotetextcopyrightpermission[1]{}

\setcitestyle{numbers,sort&compress}

\usepackage{hyperref}

\usepackage{fancyhdr}

\usepackage{amssymb,amsmath}
\usepackage{graphicx}

\usepackage{xcolor}

\usepackage{listings}

\usepackage{multirow}

\usepackage{braket}

\usepackage{array}

\usepackage{enumitem}
\setlist{leftmargin=5mm}

\usepackage[font=normal,skip=3pt]{caption}
\DeclareCaptionLabelFormat{figure}{Fig.\nobreakspace\thefigure}
\usepackage{balance}

\definecolor{feedlinegray}{gray}{0.8}
\definecolor{codegray}{gray}{0.9}
\newcommand{\code}[1]{{\small \texttt{#1}}}

\newcommand{\gate}[1]{$#1$}

\newcommand{\bin}[1]{{`\texttt{#1}'}}
\newcommand{\bits}[1]{$#1~\mathrm{bits}$}

\newcommand{\tablescript}[1]{\scriptsize{\texttt{#1}}}

\definecolor{purple}{RGB}{255, 0, 255}
\definecolor{qing}{RGB}{32, 101, 159}

\definecolor{opcodecolor}{RGB}{0,0,0}
\definecolor{eclipseBlue}{RGB}{42,0.0,255}
\definecolor{eclipseGreen}{RGB}{63,127,95}
\definecolor{eclipsePurple}{RGB}{127,0,85}
\definecolor{eclipseRed}{RGB}{223,26,65}
\definecolor{sublimegreen}{RGB}{112,200,0}
\definecolor{purple}{RGB}{255, 0, 255}

\newcommand{\ns}[1]{$#1~\mathrm{ns}$}
\newcommand{\us}[1]{$#1~\mu\mathrm{s}$}

\newcommand{\MHz}[1]{$#1~\mathrm{MHz}$}
\newcommand{\GSps}[1]{$#1~\mathrm{GSa/s}$}
\newcommand{\Tone}{T_1}
\newcommand{\cnot}{\textrm{CNOT}}

\DeclareFontFamily{U}{matha}{\hyphenchar\font45}
\DeclareFontShape{U}{matha}{m}{n}{
      <5> <6> <7> <8> <9> <10> gen * matha
      <10.95> matha10 <12> <14.4> <17.28> <20.74> <24.88> matha12
      }{}
\DeclareSymbolFont{matha}{U}{matha}{m}{n}
\DeclareMathSymbol{\wedge}         {2}{matha}{"5E}
\DeclareMathSymbol{\vee}           {2}{matha}{"5F}

\usepackage{listings}

\lstdefinelanguage{eQASM}
{
  sensitive=false, 
  alsoletter={.1234567890},
  keywords={},
  otherkeywords={ 
  |
  },
  morekeywords=[2]{always, never, eq, ne, eqz, nez, lt, ltz, le, gt, ge, gez, ltu, leu, gtu, geu, carry, notcarry},
  morekeywords=[3]{qwait, qwaitr,  SMIS, SMIT},
  morekeywords=[4]{bs},
  morekeywords=[5]{MeasZ, X, Y, CNOT, X180, Xm180, X90, Xm90, Y180, Ym180, Ym90, Y90, CZ, H, S, Z, QNOP, q_op3, q_op2, q_op1, q_op0, C_X},
  morekeywords=[6]{nop, stop, ldi, ldui, add, sub, addc, subc, and, or, xor, not, fmr, br, cmp, fbr,  st, ld},
  morekeywords=[7]{.register, .def_sym},
  morekeywords=[8]{nand, nor, xnor, bra, brn, beq, bne, blt, ble, bgt, bge, bltu, bleu, bgtu, bgeu, goto, mov, shl1, mult2},
  morecomment=[l]{\#}, 
}

\lstdefinestyle{eQASMStyle}{
  language=eQASM,
  linewidth=0.97\columnwidth,
  xleftmargin=0.06\columnwidth,
  basicstyle=\linespread{1.2}\small\ttfamily, 
  captionpos=b, 
  extendedchars=true, 
  tabsize=2, 
  columns=fixed, 
  keepspaces=true, 
  breaklines=true, 
  frame=trbl, 
  frameround=tttt, 
  numbers=left, 
  commentstyle=\color{eclipseGreen}, 
  keywordstyle=\color{eclipseGreen}, 
  keywordstyle=[2]*\bfseries\color{black}, 
  keywordstyle=[3]*\color{qing}, 
  keywordstyle=[4]\color{grayosix},
  keywordstyle=[5]*\bfseries\color{orange}, 
  keywordstyle=[6]*\color{eclipseBlue}, 
  keywordstyle=[7]\color{eclipsePurple}, 
  keywordstyle=[8]*\color{eclipseRed}
}

\lstnewenvironment{eQASM}%
{\lstset{style=eQASMStyle}}
{}

\usepackage{tikz,colortbl}
\usetikzlibrary{calc}
\usepackage{zref-savepos}

\newcounter{NoTableEntry}
\renewcommand*{\theNoTableEntry}{NTE-\the\value{NoTableEntry}}

\newcommand*{\notableentry}{%
  \multicolumn{1}{@{}c@{}|}{%
    \stepcounter{NoTableEntry}%
    \vadjust pre{\zsavepos{\theNoTableEntry t}}
    \vadjust{\zsavepos{\theNoTableEntry b}}
    \zsavepos{\theNoTableEntry l}
    \hspace{0pt plus 1filll}%
    \zsavepos{\theNoTableEntry r}
    \tikz[overlay]{%
      \draw[red]
        let
          \n{llx}={\zposx{\theNoTableEntry l}sp-\zposx{\theNoTableEntry r}sp},
          \n{urx}={0},
          \n{lly}={\zposy{\theNoTableEntry b}sp-\zposy{\theNoTableEntry r}sp},
          \n{ury}={\zposy{\theNoTableEntry t}sp-\zposy{\theNoTableEntry r}sp}
        in
        (\n{llx}, \n{lly}) -- (\n{urx}, \n{ury})
        (\n{llx}, \n{ury}) -- (\n{urx}, \n{lly})
      ;
    }%
  }%
}

\usepackage[absolute,overlay]{textpos}

\title{eQASM: An Executable Quantum Instruction Set Architecture}
\author{X.~Fu\texorpdfstring{\textsuperscript{1, 2, $\ast$}}{},~~L.~Riesebos\texorpdfstring{\textsuperscript{1, 2}}{},~~M.~A.~Rol\texorpdfstring{\textsuperscript{1,3}}{},~~J.~van~Straten\texorpdfstring{\textsuperscript{4}}{},~~J.~van~Someren\texorpdfstring{\textsuperscript{1,2}}{},~~N.~Khammassi\texorpdfstring{\textsuperscript{1,2}}{},
I.~Ashraf\texorpdfstring{\textsuperscript{1,2}}{},~~R.~F.~L.~Vermeulen\texorpdfstring{\textsuperscript{1,3}}{},~~V.~Newsum\texorpdfstring{\textsuperscript{5,1}}{},~~K.~K.~L.~Loh\texorpdfstring{\textsuperscript{5,1}}{},~~J.~C.~de~Sterke\texorpdfstring{\textsuperscript{6,1}}{},\\W.~J.~Vlothuizen\texorpdfstring{\textsuperscript{5,1}}{},~~R.~N.~Schouten\texorpdfstring{\textsuperscript{1,3}}{},~~C.~G.~Almudever\texorpdfstring{\textsuperscript{1,2}}{},~~L.~DiCarlo\texorpdfstring{\textsuperscript{1,3}}{},~~K.~Bertels\texorpdfstring{\textsuperscript{1,2$\dagger$}}{}
}
\affiliation{
  \institution{\textsuperscript{1} \textit{QuTech, Delft University of Technology, P.O. Box 5046, 2600 GA Delft, The Netherlands}\\
\textsuperscript{2} \textit{Quantum Computer Architecture Lab, Delft University of Technology, Mekelweg 4, 2628 CD Delft, The Netherlands}\\
\textsuperscript{3} \textit{Kavli Institute of Nanoscience, Delft University of Technology, P.O. Box 5046, 2600 GA Delft, The Netherlands}\\
\textsuperscript{4} \textit{Computer Engineering Lab, Delft University of Technology, Mekelweg 4, 2628 CD Delft, The Netherlands}\\
\textsuperscript{5} \textit{Netherlands Organisation for Applied Scientific Research (TNO), P.O. Box 155, 2600 AD Delft, The Netherlands}\\
\textsuperscript{6} \textit{Topic Embedded Systems B.V., P.O. Box 440, 5680 AK Best, The Netherlands}}
\{$\ast$~x.fu-1,~~$\dagger$~k.l.m.bertels\}@tudelft.nl
}
\begin{document}

\begin{textblock*}{17cm}(2.8cm, 1.5cm) 
\large \textit{In Proceedings of the 25th International Symposium on High-Performance Computer Architecture (HPCA'19)}
\end{textblock*}

\begin{abstract}
A widely-used quantum programming paradigm comprises of both the data flow and control flow. Existing quantum hardware cannot well support the control flow, significantly limiting the range of quantum software executable on the hardware. By analyzing the constraints in the control microarchitecture, we found that existing quantum assembly languages are either too high-level or too restricted to support comprehensive flow control on the hardware. Also, as observed with the quantum microinstruction set QuMIS~\cite{fu2017experimental}, the quantum instruction set architecture (QISA) design may suffer from limited scalability and flexibility because of microarchitectural constraints. It is an open challenge to design a scalable and flexible QISA which provides a comprehensive abstraction of the quantum hardware.

In this paper, we propose an executable QISA, called eQASM, that can be translated from quantum assembly language (QASM), supports comprehensive quantum program flow control, and is executed on a quantum control microarchitecture. With efficient timing specification, single-operation-multiple-qubit execution, and a very-long-instruction-word architecture, eQASM presents better scalability than QuMIS. The definition of eQASM focuses on the assembly level to be expressive. Quantum operations are configured at compile time instead of being defined at QISA design time. We instantiate eQASM into a 32-bit instruction set targeting a seven-qubit superconducting quantum processor. We validate our design by performing several experiments on a two-qubit quantum processor.
\end{abstract}
\maketitle

\pagestyle{plain}

\section{Introduction}
Quantum computing is promising with the potential to accelerate solving some problems which are inefficiently solved by classical computers, such as quantum chemistry simulation~\cite{feynman1982simulating, kassal2011simulating}. A near-term goal is to develop a fully programmable quantum computer based on the circuit model with Noisy Intermediate-Scale Quantum (NISQ) technology~\cite{preskill2018quantum} without quantum error correction~\cite{terhal2015quantum}. 
To this end, a viable way for both quantum software and hardware is to support the widely-used ``quantum data, classical control'' programming paradigm~\cite{selinger2004towards}. In this paradigm, the data flow is the state evolution subject to a sequence of classical or quantum operations, and the control flow is the order of operations that are executed, which could be directed by qubit measurement results. It is embodied in a wide range of quantum applications, including active qubit reset~\cite{riste2012feedback}, teleportation~\cite{bennett1993teleporting}, non-deterministic quantum gate decomposition~\cite{paetznick2014repeat}, iterative quantum phase estimation~\cite{kitaev1996quantum}, etc. 

Though high-level quantum software can support this paradigm well, current quantum hardware cannot because of limited executability of existing low-level quantum assembly languages. Assembly languages are introduced as human-readable representation of the interface to hardware (or machine code). But existing quantum assembly languages either incorporate too high-level constructs to be directly implemented by a microarchitecture (including QASM-HL~\cite{javadiabhari2015scaffcc}, Quil~\cite{smith2016practical}, f-QASM~\cite{liu2017qsi}, etc.), or are too restricted to provide a comprehensive abstraction of the quantum hardware which can support the required flow control (including OpenQASM~\cite{cross2017open}, QuMIS~\cite{fu2017experimental}, etc.). This fact significantly limits the range of quantum software which can be executed by the hardware. 

Required is a scalable and flexible interface which can be executed by the hardware to support the ``quantum data, classical control'' programming paradigm. 
Considering the constraints in microarchitecture implementation, designing such an interface is challenged by the difficulty of (1) providing a comprehensive abstraction of the quantum hardware~\cite{chong2017programming} and (2) making the quantum instruction set architecture (QISA) scalable and flexible.

\subsection{Comprehensive Abstraction Challenge}
Existing quantum software, including quantum programming languages~\cite{ying2016foundations}, compilers~\cite{javadiabhari2015scaffcc, wecker2014liqui, Steiger2016projectq}, and quantum assembly languages~\cite{javadiabhari2015scaffcc, smith2016practical, cross2017open, liu2017qsi}, can well describe both the data flow and the control flow. The basic constructs of control flow include procedure, selection, loop, and recursion~\cite{selinger2004towards, ying2016foundations}. Among them, selection and loop may use feedback based on qubit measurement results to select which instructions to execute in the following. 
However, existing programmable quantum hardware mostly focuses on supporting the data flow. They cannot well support the control flow because they lack programmable feedback with sufficient flexibility~\cite{ibmq, rigettiforest, alibabaquantumcloud} (though feedback has been demonstrated with customized hardware in multiple experiments~\cite{riste2012feedback, bultink2016active, ofek2016extending, hu2018demonstration}).

The difficulty of supporting programmable feedback in the hardware roots in the strict requirements on the electrical signals (e.g., precise parameters and timing) used to control qubits. To satisfy these requirements, a three-step procedure is usually used: (i) defining waveforms in digital format that are long enough to include all operations of the quantum application, (ii) uploading the digital waveforms to arbitrary waveform generators (AWG), and (iii) converting these digital waveforms into analog ones at runtime by the AWGs. Because of the computational complexity and communication latency, step (i) and (ii) are performed at static time. The fact that waveforms are determined at static time makes runtime feedback almost impossible.

Using digital waveforms only as the interface to quantum hardware is cumbersome and presents poor scalability and flexibility. To address this issue, our previous work~\cite{fu2017experimental} proposed the quantum control microarchitecture QuMA, which implements the instruction-based waveform generation as an alternative. In this method, a set of short pulses in place of long waveforms are uploaded to AWGs, with each pulse representing a quantum operation. By executing instructions in the quantum microinstruction set QuMIS, desired pulses are selected and triggered at runtime, which in principle provides the foundation for runtime feedback. 
However, the timing of executing instructions is decoupled from that of pulse generation by a set of FIFOs~\cite{fu2017experimental}. 
As a consequence, a classical instruction that uses a qubit measurement result may start execution before the expected result is ready, or read another result when there are multiple instructions measuring the same qubit, which can lead to a wrong execution result.
Hence, it is a challenge to design a mechanism that can correctly implement runtime feedback to support comprehensive quantum program flow control.

\subsection{Scalability \& Flexibility Challenges}
A scalability issue in QISA design was first observed with QuMIS, which was implemented by the control microarchitecture QuMA. 
Because quantum operations of a quantum application are applied on qubits with particular timing, quantum instructions should be fetched from the instruction memory and processed in time to ensure the described operations can be applied with the correct timing. Since every instruction can only encode a limited number of quantum operations, it would require a minimum number ($R_\mathrm{req}$) of instructions to be issued and processed per cycle for this application. 
However, the microarchitecture can only issue a limited number ($R_{\mathrm{allowed}}$) of instructions per cycle given a limited instruction issue rate. As the number of qubits grows, $R_\mathrm{req}$ in general increases. When $R_\mathrm{req} > R_\mathrm{allowed}$,  the microarchitecture cannot execute the quantum program correctly. We call this problem the \textit{quantum operation issue rate problem}~\cite{fu2017experimental, fu2018microarchitecture, tannu2017taming}.
The low instruction information density of QuMIS contributes to a large $R_\mathrm{req}$, exaggerating this problem. 
A QuMIS program has a relatively low instruction information density because (1) an explicit waiting instruction is required to separate any two consecutive timing points; (2) each target qubit of a quantum operation occupies a field in the instruction, making the instruction width a limitation for the number of target qubits in a single instruction; (3) two parallel and different operations cannot be combined into a single instruction.
The low instruction information density of QuMIS contributes to a large $R_\mathrm{req}$. 
In our previous experiments, we found that the boundary condition $R_\mathrm{req} \leq R_\mathrm{allowed}$ cannot be satisfied for some applications with only two qubits. Hence, how to design a QISA with a high instruction information density forms a scalability challenge.

Though being able to provide flexibility in directing the program flow control at runtime, some QISA design suffers from no quantum semantics. For example, instructions of QuMIS and the Raytheon BBN APS2 instruction set~\cite{ryan2017hardware} are low level and tightly bound to the electronic hardware implementation to ensure the executability. Compared to existing quantum assembly languages, these instructions are microinstructions without explicit quantum semantics. Thus, they do not qualify as a QISA, and it is another challenge to design an executable QISA with flexible quantum semantics.

\subsection{Contributions}
In this paper, we propose an executable QISA based on QASM, named \textit{executable QASM} (eQASM). eQASM can be generated by the compiler backend from a higher-level representation, like cQASM~\cite{nkhammassi2018cqasm}. eQASM contains both quantum instructions and auxiliary classical instructions to support quantum program flow control. eQASM supports a set of discrete quantum operations to enable a microarchitecture implementation. The contributions of the paper are the following:
\begin{itemize}[leftmargin=*]
    \item \textbf{Comprehensive quantum control flow:} eQASM proposes two kinds of feedback with required microarchitectural mechanisms to implement them: \textit{fast conditional execution} for simple but fast feedback, and \textit{comprehensive feedback control} (CFC) for arbitrary user-definable feedback. Based on this, eQASM can support comprehensive program flow control required by the ``quantum data, classical control'' paradigm, and significantly broadens the range of quantum applications executable on hardware;
    \item \textbf{Operational implementation:} eQASM is a QISA framework with the definition focusing on the assembly level and the basic rules of mapping assembly to binary. It requires customized instantiation for the binary format targeting a particular platform, which allows the pursuit of flexibility and practicability in microarchitecture implementation;
    \item \textbf{Increased instruction information density:} eQASM adopts \textit{Single-Operation-Multiple-Qubit} (SOMQ) execution, a Very-Long-Instruction-Word (VLIW) architecture and a more efficient method for explicit timing specification, which can considerably alleviate the quantum operation issue rate  problem when compared to QuMIS;
    \item \textbf{Configurable QISA at compile time:} As opposed to the classical instruction set architecture (ISA) whose operations are defined at ISA design time, eQASM enables the programmer to configure allowed quantum operations at compile time, leaving ample space for compiler-based optimization.
\end{itemize}

We instantiate eQASM into a 32-bit instruction set targeting a seven-qubit superconducting quantum processor and implement it using a control microarchitecture derived from QuMA as proposed in~\cite{fu2017experimental}. We validated eQASM by performing several experiments over a two-qubit superconducting quantum processor using the implemented microarchitecture.

This paper is organized as follows. Section~\ref{sec:overview} introduces the heterogeneous quantum programming model adopted by eQASM and an overview of eQASM. The quantum instructions of eQASM with related mechanisms are explained in Section~\ref{sec:arch}. Section~\ref{sec:instantiation} describes the instantiation of eQASM targeting a seven-qubit quantum processor as well as its microarchitecture and implementation. Section~\ref{sec:experiment} shows the experiments, and Section~\ref{sec:conclusion} concludes.
\section{eQASM Overview}
\label{sec:overview}

To our understanding, it is viable to integrate quantum computing in a similar way as a GPU or an FPGA in a heterogeneous architecture.
The quantum part can be seen as an accelerator for particular classically-hard tasks. 
This section introduces the eQASM programming and compilation model, the design guidelines for eQASM, the architectural state, and an overview of instructions of eQASM.

\subsection{Programming and Compilation Model}
\begin{figure}[t!]
\centering
\includegraphics[width=\columnwidth]{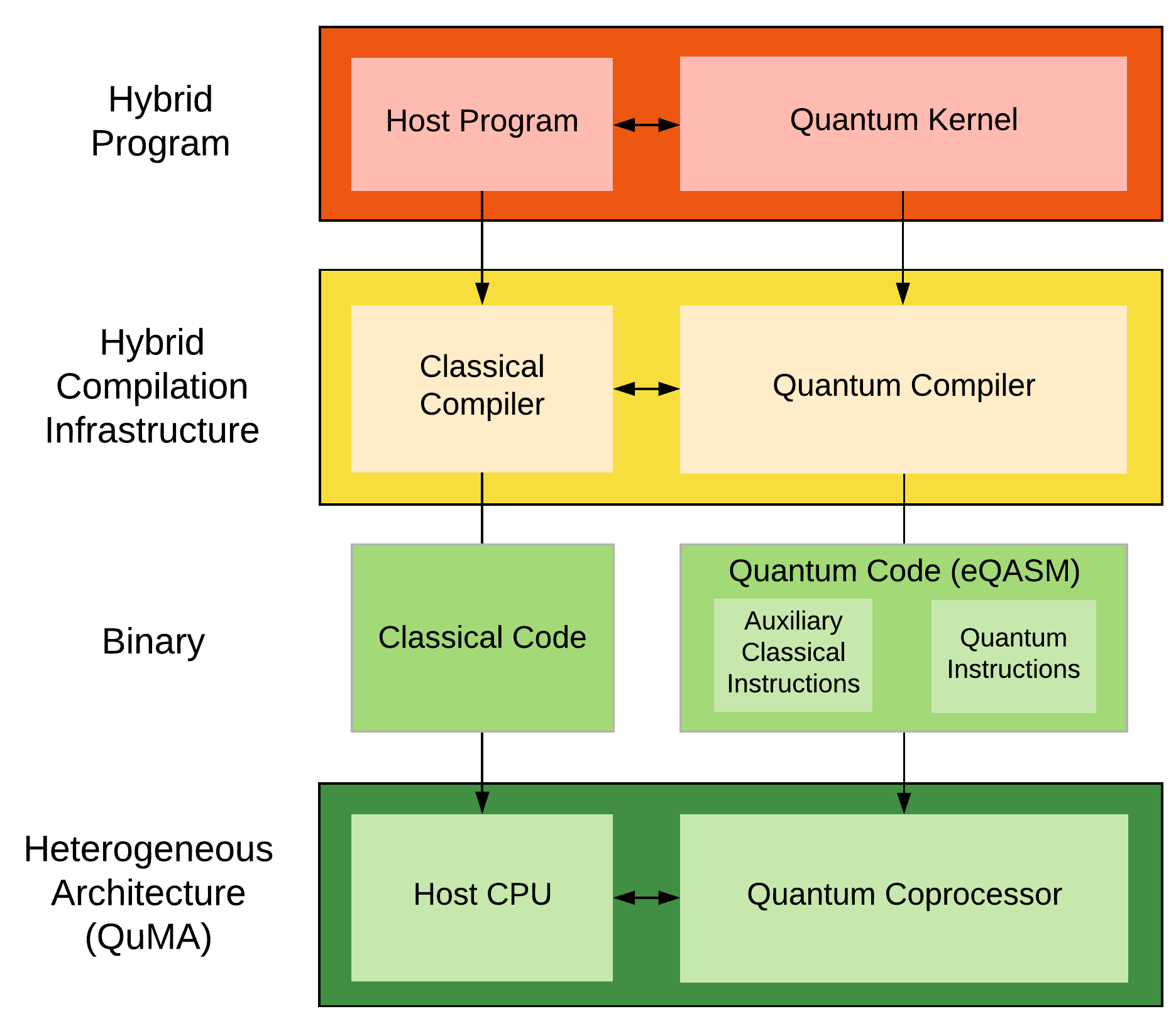}
\caption{Heterogeneous quantum programming and compilation model.}
\label{fig:prog_model}
\end{figure}

To take advantage of both classical computing and quantum computing, 
eQASM is defined based on OpenCL~\cite{stone2010opencl}, 
which is an open industry standard for classical heterogeneous parallel computing. 
The programming and compilation model of eQASM is shown in Fig.~\ref{fig:prog_model}.

A quantum-classical hybrid program contains a host program and one or more quantum kernels with the quantum kernel(s) accelerating particular parts of the computation.
The host program is described using a classical programming language, such as Python or C++, 
and the quantum kernels are described using a quantum programming language, such as Scaffold~\cite{abhari2012scaffold} or Q\#~\cite{svore2018q}.
A hybrid compilation infrastructure compiles the host program into classical code using a conventional compiler such as GCC, which is later executed by the classical host CPU.

The quantum compiler, such as OpenQL~\cite{fu2017experimental}, compiles the quantum kernels in two steps.
First, quantum kernels are compiled into QASM, or a similar format mathematically equivalent to the circuit model.
This format is hardware independent, which enables high-level optimizations and can be ported across different platforms.
Most of the hardware constraints are taken into account in the second step, 
where the compiler performs qubit mapping and scheduling, and low-level optimization. 
The output is the analog pulse configuration for physical operations~\cite{mckay2018qiskit}, the microcode defining QISA-level operations using the pulse-defined physical operations, and the quantum code consisting of eQASM instructions.
The quantum code contains quantum instructions as well as auxiliary classical instructions to support comprehensive quantum program flow control including runtime feedback~\cite{selinger2004towards, ying2016foundations}.
After the host CPU has loaded the quantum code, microcode, and pulses into the quantum processor, the quantum code can be directly executed.
In the rest of this paper, we focus on the quantum processor, i.e., the microarchitecture in charge of controlling qubits.
The interaction between the classical processor and the quantum processor is a research topic outside the scope of this paper.

\subsection{Design Guidelines}
\label{sec:guideline}
The core requirement of eQASM is being executable on real hardware but not bound to a particular electronic control setup. The design of eQASM focuses on providing an comprehensive abstraction at the architecture level which can support the ``quantum data, classical control'' programming paradigm as well as some quantum experiments such as measuring the relaxation time of qubits ($\Tone$ experiment).
Also, because some experiments and radical compiler-based optimization techniques such as quantum optimal control~\cite{werschnik2007quantum, leung17speedup} may use uncalibrated or uncommon quantum operations, eQASM should support the usage of user-defined quantum operations.
In stark contrast to classical computation where time is irrelevant to correctness, timing plays a key role in the control over qubits, i.e., in the execution of algorithms and experiments.
To ensure repeatability of quantum algorithms and experiments and reduce the risk of bug fixes or updates in software or hardware where timing is critical, timing of operations can be exposed at the architectural level as suggested by~\cite{lee2009computing}.
The design of eQASM is guided by five main principles:
\begin{enumerate}[leftmargin=*]
    \item eQASM should include classical instructions to support comprehensive quantum program flow control including runtime feedback;
    \item eQASM should contain well-defined methods to specify the timing of quantum operations;
    \item eQASM should be simple to allow a straightforward microarchitecture implementation;
    \item Low-level hardware information should be abstracted away from the eQASM assembly as much as possible to avoid eQASM being stuck to a particular hardware implementation;
    \item The quantum operation issue rate is a potential bottleneck of the quantum microarchitecture, and should be addressed, e.g., by densely encoding the instructions such as done with SIMD and VLIW for classical architectures;
    \item eQASM should be flexible to allow different quantum operations via configuration.

\end{enumerate}

\subsection{Architectural State}
\label{sec:sys_state}
As shown in Fig.~\ref{fig:arch_state}, the architectural state of the quantum processor includes:

\begin{figure}[bt]
\centering
\includegraphics[width=0.8\columnwidth]{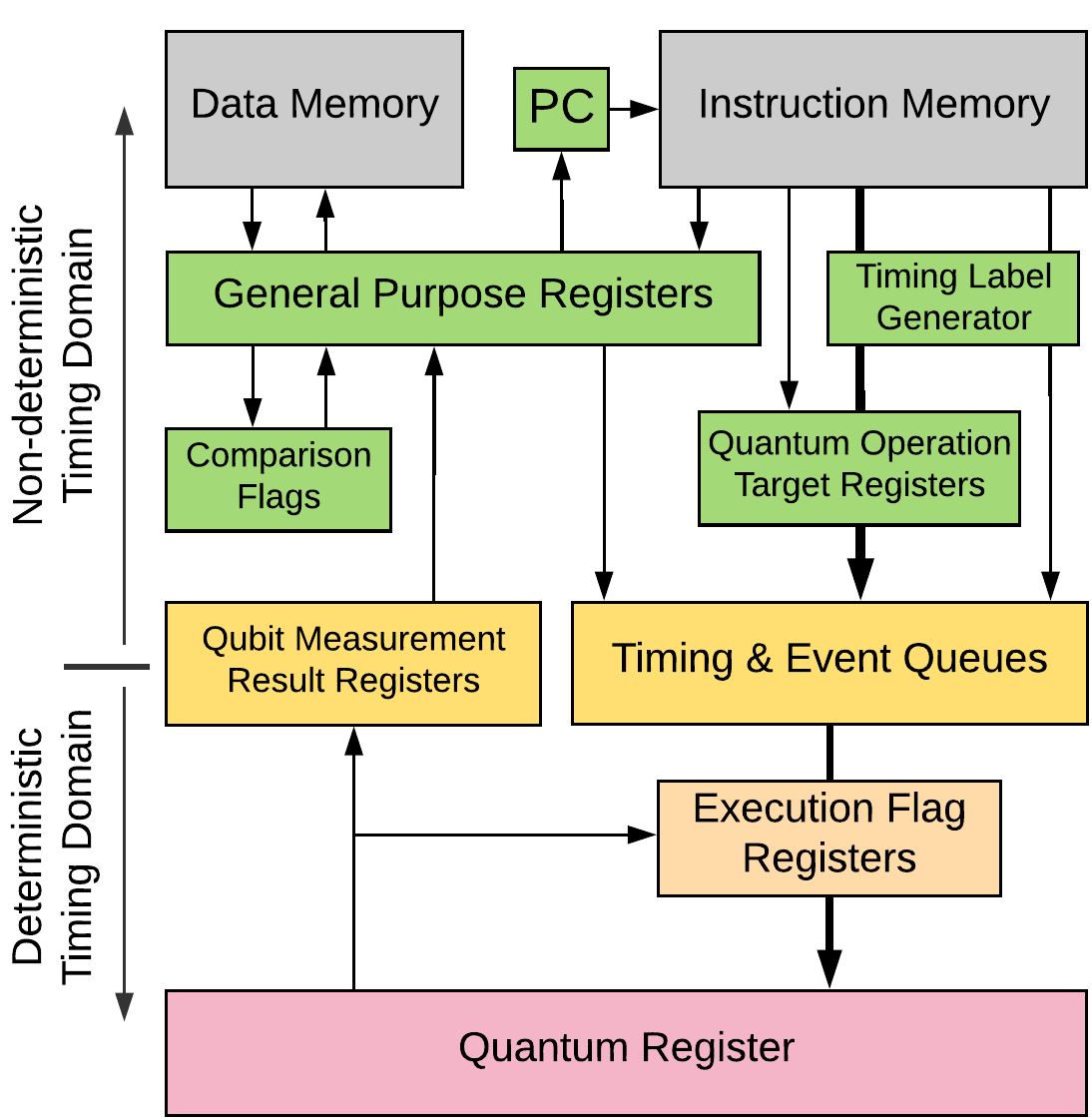}
\caption{Architectural state of eQASM. 
Arrows indicates the possible information flow.
The thick arrows represent quantum operations, which read information from the modules passed through.}
\label{fig:arch_state}
\end{figure}

\subsubsection{Data Memory}
The data memory can buffer intermediate computation results and serve as the communication channel between the host CPU and the quantum processor.

\subsubsection{Instruction Memory \& Program Counter}
The eQASM instructions are stored in the instruction memory, and the Program Counter (PC) contains the address of the next eQASM instruction to fetch.
eQASM does not define an instruction memory size or a memory hierarchy.

\subsubsection{General Purpose Registers}
The general purpose register (GPR) file is a set of 32-bit registers, labeled as \code{Ri}, where \code{i} is the register address.

\subsubsection{Comparison Flags}\label{sec:comp_flag_reg_file}
The comparison flags store the comparison result of two general purpose registers which are used by comparison and branch related instructions (see Table~\ref{tab:eqasm_insn}).

\subsubsection{Quantum Operation Target Registers}
Each quantum operation target register can be used as an operand of a quantum operation.
Since most quantum technologies support physical operations applied on up to two qubits, there are two types of quantum operation target registers: single-qubit target registers for single-qubit operations [including measurement (MEASZ)], and two-qubit target registers for two-qubit operations.

Each single- (two-)qubit target register can store the physical addresses of a set of qubits (allowed qubit pairs).
An \textit{allowed qubit pair} is a pair of qubits on which we can directly apply a physical two-qubit gate.
A single- (two-)qubit target register is labelled as \code{Si} (\code{Ti}), with \code{i} being the register address.
eQASM does not define the format of target registers (see Section~\ref{sec:somq} for a discussion).

\subsubsection{Timing and Event Queues}
To support explicit timing specification of quantum operations, 
eQASM adopts a queue-based timing control scheme~\cite{fu2017experimental}.
The timing and event queues are used to buffer timing points and operations generated from the execution of quantum instructions (see Section~\ref{sec:timing-model}).
Together with the qubit measurement result registers, 
it separates the processor into two timing domains, 
the deterministic one and the non-deterministic one.

\subsubsection{Qubit Measurement Result Registers}
Each qubit measurement result register is 1-bit wide, 
and stores the result of the last finished measurement instruction on the corresponding qubit when it is valid (see Section~\ref{sec:cfc}).
It is labeled as \code{Qi}, where \code{i} is the physical address of the qubit.

\subsubsection{Execution Flag Registers}
Sometimes, the execution of a quantum operation depends on a simple combination of previous measurement results of this qubit~\cite{riste2012feedback, bultink2016active}. To this end, each qubit is associated with an execution flag register, which contains multiple flags derived automatically by the microarchitecture from the last measurement results of this qubit.
The execution flag register file is used for fast conditional execution (see Section~\ref{sec:ffc}).

\subsubsection{Quantum Register}
The quantum register is the collection of all physical qubits inside the quantum processor.
Each qubit is assigned a unique index, known as the \textit{physical address}.
Since data in qubits can be superposed, 
eQASM does \textbf{not} allow direct access to the quantum data at the instruction level.
Instead, users can measure qubits using measurement instructions and later access the results in the qubit measurement result registers.

\subsection{Instruction Overview}

\begin{table*}[bt]
\centering
\caption{Overview of eQASM instructions.}
\label{tab:eqasm_insn}
\small
{\setlength{\extrarowheight}{2pt}%
\begin{tabular}{|c|l|l|}
\hline
\textbf{Type}      & \textbf{Syntax} & \textbf{Description} \\ \hline \hline
\multirow{2}{*}{Control}            & \lstinline!CMP  Rs, Rt! & \begin{tabular}[c]{@{}l@{}}\textbf{C}o\textbf{MP}are GPR \texttt{Rs} and \texttt{Rt} and store the result into the
                                                                        comparison flags.\end{tabular} \\ \cline{2-3}
                                    & \lstinline!BR   <Comp. Flag>, Offset! & \begin{tabular}[c]{@{}l@{}}(\textbf{BR}anch) Jump to \texttt{PC~+~Offset} if the specified comparison flag is \bin{1}.\end{tabular} \\ \hline
\multirow{6}{*}{Data Transfer} & \lstinline!FBR  <Comp. Flag>, Rd! & (\textbf{F}etch \textbf{B}ranch \textbf{R}egister) Fetch the specified comparison flag into GPR Rd. \\ \cline{2-3}
                               & \lstinline!LDI  Rd, Imm!   & (\textbf{L}oa\textbf{D} \textbf{I}mmediate) Rd = sign\_ext(Imm{[}19..0{]}, 32). \\ \cline{2-3}
                               & \lstinline!LDUI Rd, Imm, Rs! & (\textbf{L}oa\textbf{D} \textbf{U}nsigned \textbf{I}mmediate) Rd = Imm{[}14..0{]}::Rs{[}16..0{]}. \\ \cline{2-3}
                               & \lstinline!LD   Rd, Rt(Imm)! & (\textbf{L}oa\textbf{D} from memory) Load data from memory address \texttt{Rt + Imm} into GPR \texttt{Rd}.  \\ \cline{2-3}
                               & \lstinline!ST   Rs, Rt(Imm)! & (\textbf{ST}ore to memory) Store the value of GPR \texttt{Rs} in memory address \texttt{Rt + Imm}. \\ \cline{2-3}
                               & \lstinline!FMR  Rd, Qi! & \begin{tabular}[c]{@{}l@{}}(\textbf{F}etch \textbf{M}easurement \textbf{R}esult) Fetch the result of the last measurement instruction on qubit\\ \texttt{i} into GPR \texttt{Rd}.\end{tabular}\\ \hline
Logical & \begin{tabular}[c]{@{}l@{}}\lstinline!AND/OR/XOR Rd, Rs, Rt!\\ \lstinline!NOT  Rd, Rt!\end{tabular} & Logical and, or, exclusive or, not. \\ \hline
Arithmetic                     & \lstinline!ADD/SUB Rd, Rs, Rt! & Addition and subtraction. \\ \hline
\hline
Waiting & \begin{tabular}[c]{@{}l@{}} \lstinline!QWAIT   Imm!\\ \lstinline!QWAITR  Rs! \end{tabular} & \begin{tabular}[c]{@{}l@{}}(\textbf{Q}uantum \textbf{WAIT} \textbf{I}mmediate/\textbf{R}egister) Specify a timing point by waiting for the \\number of cycles indicated by the immediate value \texttt{Imm} or the value of GPR \texttt{Rs}.\end{tabular} \\ \hline
Target Specify & \begin{tabular}[c]{@{}l@{}} \lstinline!SMIS Sd, <Qubit List>! \\ \lstinline!SMIT Td, <Qubit Pair List>! \end{tabular} & \begin{tabular}[c]{@{}l@{}}(\textbf{S}et \textbf{M}ask \textbf{I}mmediate for \textbf{S}ingle-/\textbf{T}wo-qubit operations) Update the single- (two-)qubit \\operation target register \texttt{Sd} (\texttt{Td}).\end{tabular}\\ \hline
Q. Bundle &  \lstinline![PI,] Q_Op [| Q_Op]*! & Applying operations on qubits after waiting for a small number of cycles indicated by \texttt{PI}. \\ \hline
\end{tabular}}
\end{table*}

Quantum technology is evolving rapidly and is still far away from a stable state.
To avoid the format of eQASM being stuck to a specific quantum technology implementation with particular properties, the definition of eQASM focuses on the assembly level and introduces basic rules of mapping the assembly code to binary instructions.
The binary format is defined during the instantiation of eQASM targeting a concrete control electronic setup and quantum chip.
This fact enables the eQASM assembly to be expressive while leaving considerable freedom to the (micro)architecture designer to pursue microarchitectural practicability and performance.

An eQASM program can consist of interleaved quantum instructions and auxiliary classical instructions.
An overview of the eQASM instructions is shown in Table~\ref{tab:eqasm_insn}.
Since the host CPU can provide classical computation power, 
auxiliary classical instructions are simple instructions to support the execution of quantum instructions.
Complex instructions (e.g., floating-point instructions) are not included.
The top part of Table~\ref{tab:eqasm_insn} contains the auxiliary classical instructions.
There are four types: \textit{control}, \textit{data transfer}, \textit{logical}, and \textit{arithmetic} instructions. 
These are all scalar instructions.
The function \code{sign\_ext(Imm, 32)} sign extends the immediate value \code{Imm} to \bits{32}.
The operator \code{::} concatenates the two bit strings.
The \lstinline!CMP! instruction sets all comparison flags based on the comparison result of GPR \code{Rs} and \code{Rt}.
The \lstinline!BR! instruction changes the PC to \lstinline!PC + Offset! if the specified comparison flag is \bin{1}.
To enable arithmetic or logical operations on the comparison result, the \lstinline!FBR! instruction fetches the specified comparison flag into GPR \code{Rd}.
The \lstinline!FMR! instruction supports comprehensive feedback control and is explained in Section~\ref{sec:cfc}.

The bottom part of Table~\ref{tab:eqasm_insn} contains the quantum instructions.
There are three types of instructions:
\begin{itemize}
    \item Waiting instructions used to specify timing points (\lstinline!QWAIT!, \mbox{\lstinline!QWAITR!),}
    \item Quantum operation target register setting instructions (\lstinline!SMIS!, \lstinline!SMIT!), and
    \item Quantum bundle instructions, which consist of the specification of a small waiting time and multiple quantum operations.
\end{itemize}

These quantum instructions have several features based on the following four observations:
\begin{itemize}[leftmargin=*]
\item[\tiny $\blacksquare$] Many quantum experiments, such as the $T_1$ experiment, require changing the timing of operations explicitly.
Also, the timing of operations can significantly impact the fidelity of the final result as quantum errors accumulate during computation (see Section~\ref{sec:experiment}).
In the NISQ era, quantum errors accumulate as the computation is going on. 
As a consequence, the timing of operations has a significant impact on the fidelity of the final computation result (see Section~\ref{sec:experiment}).
eQASM can explicitly specify the timing of quantum operations to support quantum experiments and compiler-based timing optimization.
This enables the programmer to schedule and time the quantum operations to achieve higher fidelity.
The timing model is explained in Section~\ref{sec:timing-model}.

\item[\tiny $\blacksquare$]
Different quantum experiments or algorithms may require a different set of physical quantum operations.
To allow using different sets of quantum operations, quantum operations are specified by programmers at compile time via configuration (see Section~\ref{sec:decode}) instead of being defined at QISA design time.
This flexibility reserves ample space for compiler-based optimization.
Only single- and two-qubit operations are allowed, and more-qubit operations should be decomposed into single- and two-qubit operations by the compiler~\cite{deutsch1995universality, lloyd1995almost, divincenzo1998quantum, kudrow2013quantum, wecker2014liqui}.

\item[\tiny $\blacksquare$] To alleviate the quantum operation issue rate problem, we can reduce the required instruction issue rate $R_{\mathrm{req}}$ by increasing the instruction information density at the architecture level, and/or increase the available instruction issue rate $R_{\mathrm{allowed}}$ at the microarchitecture level. At the architecture level, eQASM reduces $R_{\mathrm{req}}$ by adopting SOMQ execution, which supports applying a single quantum operation on multiple qubits (Section~\ref{sec:somq}), and a VLIW architecture, which can combine multiple different quantum operations into a quantum bundle (Section~\ref{sec:vliw}).
eQASM adopts the VLIW design because the microarchitecture implementation can be much simpler compared to a superscalar design. The tradeoff is that the compiler needs to perform more scheduling over the instructions as suggested by~\cite{heckey2015compiler}.
The microarchitecture can also introduce multiple-issue mechanisms as classical superscalar processors to increase $R_{\mathrm{allowed}}$, which is out of the scope of this paper focusing on the architecture design.

\item[\tiny $\blacksquare$]
Two kinds of feedback are supported.
Fast conditional execution performs a Go/No-go decision for every single-qubit operation based on a execution flag of the target qubit (see Section~\ref{sec:ffc}).
To be more flexible, CFC allows programmers to define arbitrary feedback by redirecting the program flow based on the measurement results (see Section~\ref{sec:cfc}).

\end{itemize}

\section{Architecture}
\label{sec:arch}
In this section, we construct the assembly syntax of quantum operations by introducing the aforementioned mechanisms.

\hypertarget{timing-model}{%
\subsection{Timing Model}\label{sec:timing-model}}
\subsubsection{Queue-based Timing Control}
eQASM adopts the queue-based timing control scheme proposed in~\cite{fu2017experimental} since it can support explicit timing specification.
We briefly introduce this scheme and refer readers to the original paper for a detailed discussion.

In the queue-based timing control scheme, the execution of quantum instructions can be divided into a \textit{reserve} phase in the non-deterministic timing domain and a \textit{trigger} phase in the deterministic timing domain.
A timeline is constructed by the reserve phase and consumed by the trigger phase:
the result of executing quantum instructions in the reserve phase is consecutively creating new timing points on the timeline and associating events to them;
the deterministic timing domain maintains a timer, and triggers all quantum operations associated with the timing point on the timeline that it reaches.
Auxiliary classical instructions and mask setting instructions are not directly associated with timing points.
The trigger phase is handled by the microarchitecture; we introduce the reserve phase in the following.

\subsubsection{Timeline Construction}
Quantum instructions fetched from the instruction memory form a quantum instruction stream.
Instructions in the stream are executed in order; this constructs a timeline by generating consecutive timing points and assigning operations to them.

If the fetched instruction is a waiting instruction, \lstinline!QWAIT Imm! or \lstinline!QWAITR Rs!, a new timing point in the timeline is generated.
The position of the new timing point is determined by the specification of the interval since the last generated timing point.
The interval length comes from the immediate value \code{Imm} or GPR \code{Rs}.
The first timing point of the timeline can be set by a dedicated instruction, or by an external trigger to the microarchitecture.
Both waiting instructions use the unit cycle for the interval length.

If the fetched instruction is a quantum bundle instruction, the quantum operation(s) specified in the bundle instruction is associated with the last generated timing point.
If multiple quantum operations are associated to the same timing point, these quantum operations will all start execution at that same timing point.

Based on our observation over some testbenches (see Section~\ref{sec:implementation}), short intervals between timing points are a common case.
To improve the quantum operation issue rate, eQASM allows merging a \lstinline!QWAIT PI! instruction followed by a quantum operation \lstinline!<Quantum Operation>!
\begin{center}
\lstinline!QWAIT PI!\\
\lstinline!<Quantum Operation>!
\end{center}
into a single instruction
\begin{center}
\lstinline![PI,] <Quantum Operation>!.
\end{center}
Square brackets \code{[$\ldots$]} indicate that the content inside is optional.
\code{PI} is short for \code{pre\_interval}, which specifies a short interval between last generated timing point and the one when the operations in this instruction are to be triggered.
It defaults to 1 if not specified.
Value 0 is acceptable to both the \code{PI} and the waiting instructions,  which means that the following timing point is identical to the last timing point.

\subsubsection{Example}
Assuming the durations of quantum operations \mbox{\lstinline!q_op0!,} \lstinline!q_op1!, \lstinline!q_op2!, and \lstinline!q_op3! all equal one-cycle time, the following code triggers these four operations back-to-back.
\begin{lstlisting}[basicstyle=\linespread{1.1}\small\ttfamily]
LDI  r0, 1     # r0 <- 1
q_op0
q_op1          # Default PI = 1
QWAITR r0      # Register-valued waiting
0, q_op2
QWAIT 0        # Equivalent to NOP
1, q_op3       # Explicity PI = 1
\end{lstlisting}

\hypertarget{sec-decode}{
\subsection{Quantum Operation Definition \& Decoding}\label{sec:decode}}
Depending on the qubit technology and the algorithm to run, different quantum operations can be used.
eQASM does not define a fixed set of quantum operations at QISA design time, such as $\{H$, $T$, $\cnot$, $\cdots\}$. Instead, the available quantum operations can be configured by the programmer at compile time.

Flexible quantum operation configuration is achieved through the configuration of the assembler, the microcode unit and the pulse generator of the microarchitecture: on the one hand, the assembler is configured to translate a quantum operation, e.g., the $X$ gate, to the expected opcode, e.g., 0x01; on the other hand, the microcode unit translates the quantum opcodes into the expected microinstruction(s) using a microcode-based instruction decoding scheme~\cite{vassiliadis2003microcode}.
Each microinstruction represents one or more micro-operations, which are finally converted into pulses by the pulse generator with precise timing applying operations on qubits. The assembler, the microcode unit, and the pulse generator should be configured consistently at compile time.

\hypertarget{sec-address-mechanism}{
\subsection{Address Mechanism}\label{sec:somq}}
A quantum operation applied on multiple qubits is a common case.
For example, quantum computation usually starts by preparing the superposition state from initialized qubits, which  requires applying Hadamard gates on multiple qubits.
eQASM uses SOMQ execution, which can apply a single quantum operation on multiple qubits at the same time.
SOMQ is similar to classical single-instruction-multiple-data (SIMD) execution~\cite{flynn1972some}, with the operation target replaced by qubits. An instantiated eQASM can also be treated as an implementation of the previously proposed Multi-SIMD($k,d$) architecture~\cite{heckey2015compiler} but removing the assumption of SIMD regions that in each region only a single quantum operation can be applied.

SOMQ is based on an indirect qubit addressing mechanism.
The \lstinline!SMIS! or \lstinline!SMIT! instruction first defines a set of quantum operation target(s) in a quantum operation target register.
Then a quantum operation can use the target register as the operand:
\begin{center}
\lstinline!<Operation Name>  <Target Register>!.
\end{center}

\subsubsection{Address of Allowed Qubit Pairs}
Since a two-qubit operation, such as a \lstinline!CNOT! gate, can operate on its qubits differently, two qubits with different orders, i.e., \mbox{\lstinline!(Qubit A, !}\allowbreak\lstinline!Qubit B)! and \mbox{\lstinline!(Qubit B, !}\allowbreak\lstinline!Qubit A)!, are treated as different allowed qubit pairs.
The term \textit{quantum chip topology} indicates the available qubits and allowed qubit pairs of a quantum chip (see Fig.~\ref{fig:topo} for an example).
The quantum chip topology can be represented as a graph where each available qubit can be denoted as a vertex, and an allowed qubit pair as a directed edge.
In the directed edge \lstinline!(Qubit A, Qubit B)!, \lstinline!Qubit A!  is called the source qubit and \lstinline!Qubit B! the target qubit of the pair.

\subsubsection{Translation from Assembly to Binary}
Since the efficiency of encoding the qubit list (qubit pair list) may depend on the target quantum chip topology, the designer can choose different binary encoding schemes for different target quantum processors during eQASM instantiation.
In general, it is more efficient to put the address pairs in the instruction for a highly-connected quantum processor, while a mask format could be more efficient when the qubit connectivity is limited.
For example, since at most two two-qubit gates can be applied and each qubit can be addressed with \bits{3} in a fully connected 5-qubit trapped ion processor~\cite{debnath2016demonstration}, only $2\times 2\times 3~\mathrm{bits}=12~\mathrm{bits}$ are required to specify the target of a two-qubit gate. This is more efficient than a mask of $20~\mathrm{bits}$ with each bit in the mask indicating one of all 20 different allowed qubit pairs selected or not. In contrast, a mask of $6~\mathrm{bits}$ is more efficient for the IBM QX2~\cite{ibmqx2}, which also contains five qubits but has only six allowed qubit pairs.

\subsubsection{Example}
The following code sets the single-qubit target register \code{S7} to contain two qubits (0 and 1), and then applies an \gate{X} gate on both qubits simultaneously.
\begin{lstlisting}[basicstyle=\linespread{1.1}\small\ttfamily]
SMIS S7, {0, 1}
Y    S7
\end{lstlisting}

The following code sets the two-qubit target register \code{T3} to contain two pairs of qubits (1, 3) and (2, 4), and then applies a  \lstinline!CNOT! gate on them.
\begin{lstlisting}[basicstyle=\linespread{1.1}\small\ttfamily]
SMIT T3, {(1, 3), (2, 4)}
CNOT T3
\end{lstlisting}

\hypertarget{sec-vliw}{%
\subsection{Very Long Instruction Word}\label{sec:vliw}}
\subsubsection{Quantum Bundle Format}
Apart from SOMQ, different operations are also allowed to be applied on different qubits in parallel.
eQASM can combine parallel quantum operations into a \textit{quantum bundle} in a VLIW format.
We define parallel quantum operations as operations starting at the same timing point, regardless of the duration of each operation.
The format of a quantum bundle is:
\begin{center}
\lstinline![PI,] <Quantum Operation> [| <Quantum Operation>]*!
\end{center}
The vertical bar \code{|} is used to separate different quantum operations in the same bundle.
The asterisk \code{*} means the item in square brackets can repeat for $n\ge0$ times.

\subsubsection{Translation from Assembly to Binary}
In the assembly code, an arbitrary number of quantum operations can be combined into a single quantum bundle.
However, a single instruction can accommodate only a few quantum operations because of the limited instruction width.
The \textit{VLIW width} of eQASM characterizes the number of quantum operations that can be put in a single instruction word, which is defined during eQASM instantiation.
Matching this, a single quantum bundle can be broken into multiple quantum bundle instructions with \texttt{PI} being 0.
If the number of operations is not a multiple of the VLIW width, quantum no-operations (QNOP) fill up the last instruction.
For example, given a VLIW width of 2, the bundle
\begin{center}
\lstinline!PI, X S5 | H S7 | CNOT T3!
\end{center}
can be decomposed by the assembler to two consecutive quantum bundle instructions
\begin{center}
\lstinline!PI, X S5 | H S7!\\
\lstinline!0, CNOT T3 | QNOP!.
\end{center}

\subsubsection{Example}
In the code as shown in Fig.~\ref{fig:twoqubitallxy}, the instruction \lstinline!QWAIT 10000! initializes both qubits by idling them for \us{200} (assuming a cycle time of \ns{20}).
Line~6 applies a $Y$ gate on both qubits using SOMQ.
Line~7 is a VLIW instruction, which applies an $X_{90}$ and $X$ gate on each qubit.
In this paper, $X_{90}$ ($Y_{90}$) denotes the gate rotating the quantum state along the $x$- ($y$-)axis by a $\pi/2$ angle.
$X_{\mathrm{m}90}$ ($Y_{\mathrm{m}90}$) denotes similar gates but with the rotation angle of $-\pi/2$.
Line 8 measures both qubits using SOMQ.
According to the \texttt{PI} value, the $Y$ gate happens immediately after the initialization, followed by the $X_{90}$ and $X$ gates \ns{20} later and the measurement \ns{40} later.
The \us{1} waiting time (line 9) ensures no operations happening during the measurement.
\begin{figure}[hbt]
\begin{lstlisting}[basicstyle=\linespread{1.1}\small\ttfamily]
SMIS S0, {0}
SMIS S2, {2}
SMIS S7, {0, 2}
...
QWAIT    10000
0, Y     S7
1, X90   S0  | X  S2
1, MEASZ S7
QWAIT    50
...
\end{lstlisting}
\caption{Part of the code for a two-qubit \textit{AllXY} experiment, which is used in validating eQASM in Section~\ref{sec:experiment}.}
\label{fig:twoqubitallxy}
\end{figure}

\hypertarget{sec-fast-conditional-execution}{
\subsection{Fast Conditional Execution}\label{sec:ffc}}
Fast conditional execution allows executing or canceling a single-qubit operation when the micro-operation is triggered.
The decision is made based on the value of a selected flag in the execution flag register corresponding to the target qubit.
The value of the execution flag is derived by the microarchitecture using predefined combinatorial logic from the last measurement results of the same qubit.
Once there returns a measurement result for a qubit, the corresponding execution flags are updated automatically.
If the execution flag is \bin{1}, then the operation executes; otherwise, it is canceled.
A selection signal is required for each micro-operation to select which execution flag to use, which can be generated by the microcode unit, or specified by an instruction field~\cite{balensiefer2005evaluation}.
Except for the default execution flag that should always be \bin{1}, which and how many execution flags there are, should be defined during eQASM instantiation (see Section~\ref{sec:uarch} for an example).

\subsubsection*{Example}
In one instantiation of eQASM, the quantum operation \lstinline!C_X! uses the execution flag which is \bin{1} if and only if (\textbf{iff}) the last measurement result of the qubit is $\ket{1}$.
Figure~\ref{fig:ffc} shows the code for the active qubit reset experiment, where qubit 2 is put in an equal superposition using an $X_{90}$ gate after initializing it in the $\ket{0}$ state by idling it for \us{200}.
After a measurement, a conditional \lstinline!C_X! gate is applied to reset the qubit.
Qubit 2 is measured again to read out the final state for verification.
\\
\begin{figure}[hbt]
\begin{lstlisting}[basicstyle=\linespread{1.1}\small\ttfamily]
SMIS   S2, {2}
QWAIT  10000
X90    S2
MEASZ  S2
QWAIT  50
C_X    S2
MEASZ  S2
\end{lstlisting}
\caption{eQASM program for active qubit reset.
This experimental result is shown in Section~\ref{sec:experiment}.}
\label{fig:ffc}
\end{figure}

\subsection{Comprehensive Feedback Control}
\label{sec:cfc}
CFC allows adjusting the program flow based on measurement results of any qubits to enable arbitrary user-defined feedback.
This flexibility comes at the cost of longer feedback latency.
We propose a three-step mechanism to implement CFC:
\begin{enumerate}
    \item A measurement instruction is applied on the condition qubit \code{i}.
    At the moment that this measurement instruction is issued, \code{Qi} is invalidated.
    At the moment the measurement result is available, it is written in \code{Qi}.
    \code{Qi} turns back to valid if there are no more pending measurement instructions on qubit \code{i}.
    \item The \lstinline!FMR Rd, Qi! instruction fetches the value of the quantum measurement result register \code{Qi} into GPR \code{Rd}.
    If \code{Qi} is invalid, \lstinline!FMR! should wait until \code{Qi} gets valid again.
    Thereafter, the value of \code{Qi} can be fetched into \code{Rd}.
    \code{Qi} remains valid until qubit \code{i} is measured again.
    \item GPR \code{Rd} is then used in a \lstinline!BR! instruction to select the program flow to follow.
    Note, multiple \lstinline!FMR! and \lstinline!BR! instructions can be combined to support more complex feedback logic.
\end{enumerate}

\subsubsection*{Example}
The eQASM program shown in Fig.~\ref{fig:cfc} first measures qubit 1.
If the measurement result is 1, a \gate{Y} gate is applied on qubit 0, otherwise, an \gate{X} gate is applied.

\begin{figure}[hbt]
\begin{lstlisting}[basicstyle=\linespread{1.1}\small\ttfamily]
  SMIS  S0, {0}
  SMIS  S1, {1}
  LDI   R0, 1
  MeasZ S1
  QWAIT 30
  FMR   R1, Q1      # fetch msmt result
  CMP   R1, R0      # compare
  BR    EQ, eq_path # jump if R0 == R1
ne_path:
  X     S0    # happen if msmt result is 0
  BR  ALWAYS, next # this flag is always `1'
eq_path:
  Y     S0    # happen if msmt result is 1
next:
  ...
\end{lstlisting}
\caption{eQASM program using CFC.}
\label{fig:cfc}
\end{figure}
\section{Instantiation \& Implementation}
\label{sec:instantiation}
This section introduces an instantiation, microarchitecture, and implementation of eQASM.

\subsection{Target Superconducting Quantum Chip}
\begin{figure}[bt]
\centering
\includegraphics[width=0.9\columnwidth]{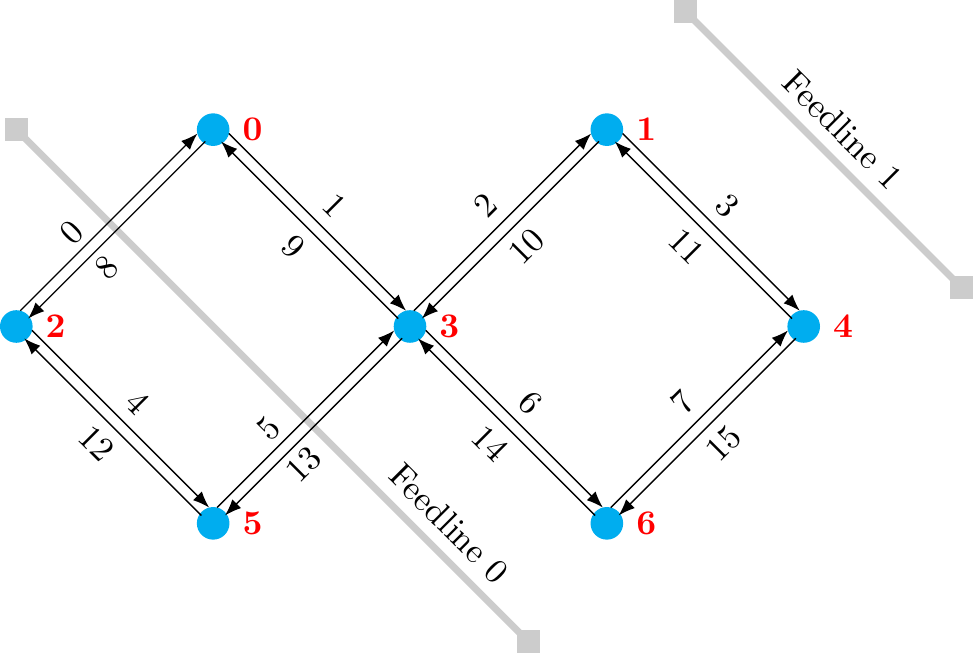}
\caption{Quantum chip topology of the target seven-qubit superconducting quantum chip.
Numbers in red are the physical addresses of qubits.
The numbers along the direct edges are addresses of the allowed qubit pairs.}
\label{fig:topo}
\end{figure}

The quantum chip topology of the target seven-qubit superconducting quantum chip is shown in Fig.~\ref{fig:topo}.
It is part of a two-dimensional square lattice as proposed in~\cite{versluis2017scalable}.
It can implement a distance-2 surface code~\cite{fowler2012surface}, which can detect one physical error.
In this figure, a vertex represents a qubit, and a directed edge represents an allowed qubit pair.
Numbers besides the vertex (edge) are the addresses of qubits (allowed qubit pairs).
For example, allowed qubit pair 0 has qubit 2 as the source qubit and qubit 0 as the target qubit.
The feedlines are used to measure the nearby coupled qubits.
Qubit 0, 2, 3, 5, and 6 (1 and 4) are coupled to feedline~0 (1).
Each feedline has an input port and an output port.
Besides, each qubit is connected to a microwave port and a flux port, which are not shown in Fig.~\ref{fig:topo}.

Operations supported by this quantum processor include measurements, single-qubit $x$- or $y$-axis rotations, and a two-qubit controlled-phase (CZ) gate.
A typical gate time is \ns{20} for single-qubit gates and \ns{\sim40} for two-qubit gates.
The duration of a measurement is typically \ns{300} - \us{1}.
A cycle time of \ns{20} is used in this instantiation.

\subsection{Instantiation Design Space Exploration}
To determine a suitable eQASM instantiation configuration for the target quantum processor [a single- (two-)qubit gate time of 1 (2) cycle(s), and a measurement time of 15 cycles], we perform analysis over three benchmarks using a quantum control architecture simulator derived from the previously proposed QPDO~\cite{riesebos2017pauli}.
Because substantial time is spent on calibrating qubits before running applications with NISQ technology, the first benchmark we select is the widely-used calibration experiment randomized benchmarking (RB)~\cite{magesan2011scalable, epstein2014investigating}, which might be limited by the high memory consumption when the required waveform for control is plainly stored in memory.
Each qubit is subject to 4096 single-qubit Clifford gates which have been decomposed into $x$ and $y$ rotations.
Because every gate happens immediately following the previous one, randomized benchmarking cannot reveal timing patterns of quantum operations in real quantum algorithms, where the parallelism is limited by two-qubit gates.
Addressing this, we also select two benchmarks from ScaffCC~\cite{javadiabhari2015scaffcc} as the representatives of small-scale quantum algorithms that might be executed with NISQ technology: a parallel algorithm (Ising model using 7 qubits, IM) which has $<1\%$ two-qubit gates, and a relatively sequential algorithm (Grover's algorithm to calculate the square root using 8 qubits, which is the minimum number of qubits required, SR), which has $\sim39\%$ two-qubit gates.
The evaluation metric is the total number of instructions.

\begin{figure}[t!]
\centering
\includegraphics[width=\columnwidth]{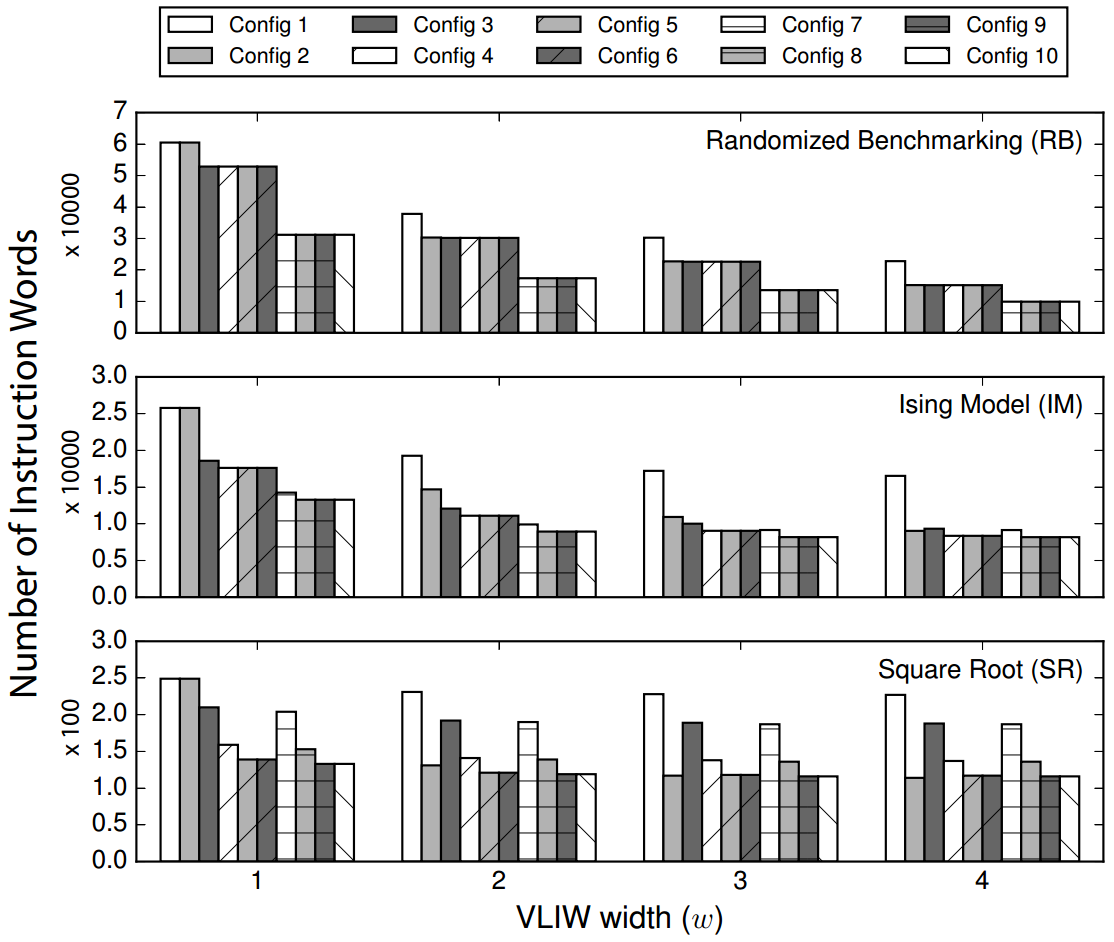}
\caption{Number of instructions for various architecture configurations for randomized benchmarking (RB), Ising model (IM), and square root (SR).}
\label{fig:arch_expl}
\end{figure}

We investigate the impact of the VLIW width ($w$), three timing-specification methods, and SOMQ on the number of instructions.
The three timing-specification methods include: the QuMIS fashion (specifying every timing point using separate \lstinline!QWAIT! instructions, $ts_1$); including \lstinline!QWAIT! in the quantum bundle instruction at the place of a quantum operation ($ts_2$); and using \code{PI} with various bit widths ($w_\mathrm{PI}$) to specify a small waiting time and using separate \lstinline!QWAIT! instructions to specify longer waiting times ($ts_3$).
The simulation results are shown in Fig.~\ref{fig:arch_expl}.

Config 1 is ($ts_1$, no \code{PI}, no SOMQ), and Config 1 with $w=1$ is chosen as the baseline.
By increasing $w$ from 1 to 4, the number of instructions can be reduced up to 62\% (RB).
Benchmarks with substantial parallelism (RB and IM) benefit more from a big $w$.
The instruction reduction in SR ($\sim8\%$) indicates that large $w$ slightly improves quantum applications with limited parallelism.

Config 2 is ($ts_2$, no \code{PI}, no SOMQ).
A minimum $w$ of 2 is required by $ts_2$ to distinguish it from $ts_1$.
Compared with Config 1, by including the \lstinline!QWAIT! operation as part of a quantum bundle instruction, Config 2 can reduce the number of instructions by 20 - 33\% (RB), 24 - 45\% (IM), 43 - 50\% (SR) by varying $w$ from 2 to 4.
SR benefits most because of two reasons.
First, due to its sequential nature, it has relatively more \lstinline!QWAIT! instructions.
Second, limited parallelism in this algorithm leaves potential VLIW slots unused, which can be filled by \lstinline!QWAIT! instructions.

Config 3/4/5/6 is ($ts_3$, $w_{\mathrm{PI}}=1/2/3/4$, no SOMQ).
Config 3 can reduce the number of instructions by 13 - 33\% for RB and 28 - 44\% for IM with $w$ varying from 1 to 4 compared with Config 1.
Since the intervals between operations in RB and IM are mostly close to 1, further increasing $w_{\mathrm{PI}}$ up to \bits{4} introduces marginal benefit.
Config 3 reduces the number of instructions of SR by $\sim17\%$ regardless of $w$.
Further increasing $w_{\mathrm{PI}}$ to $3$ or \bits{4} can reduce the number of instructions of SR by up to 48\%.
Like SR, quantum algorithms are scheduled to be executed in a time as short as possible.
This result of Config 3-6 suggests that most of the waiting time is short and can be encoded in a 3-bit \code{PI} field.
Note that Config 3/4/5 is also more beneficial than Config 2 when $w=1$ or $w=2$.

Config 7/8/9/10 is ($ts_3$, $w_{\mathrm{PI}}=1/2/3/4$, SOMQ).
Our analysis assumes that the target registers can always provide the required qubit (pair) list, and therefore shows the theoretical maximum benefit that can be obtained by SOMQ.
Compared to Config 3/4/5/6, SOMQ can introduce a maximum reduction of 42\% (Config 8, $w=2$) in the number of instructions for RB, while it can only reduce at most 4\% instructions for SR (Config 8, $w=1$).
Regardless of $w_{\mathrm{PI}}$, SOMQ can help reduce the number of instructions of IM by $\sim24$, $19$, $9$, and $2\%$ for different $w$.
This fact suggests that SOMQ is more effective for highly parallel applications, especially when $w$ is small.
An application that would benefit significantly from SOMQ is quantum error correction, which requires performing well-patterned error syndrome measurements repeatedly presenting high parallelism.
As not shown in the figure, we also analyzed the number of effective quantum operations in each quantum bundle for Config 9, which is 1.795, 2.296, and 3.144 for RB, 1.485, 1.622, and 1.623 for IM, and 1.118, 1.147, and 1.147 for SR with $w$ varying from 2 to 4, respectively.
It indicates that with the existence of SOMQ, $w>2$ is not highly required for many quantum applications (RB is a special case with extreme parallelism).

As a result of the analysis, our eQASM instantiation adopts Config 9 ($ts_3$, $w_{\mathrm{PI}}=3$, SOMQ) with $w=2$.
A width of \bits{32} is used by all instructions for the memory alignment.
Two instruction formats are used: the single format with the highest bit being \bin{0} and the bundle format with the highest bit being \bin{1}.
Single format instructions use the other \bits{31} to encode a single instruction, including all auxiliary classical instructions, and \lstinline!SMIS!, \lstinline!SMIT!, \mbox{\lstinline!QWAIT(R)!} instructions.
For brevity, we only present the format of quantum instructions as shown in Fig.~\ref{fig:bundle_fmt}.

\begin{figure}[tb]
\footnotesize
\centering
\begin{tabular}{>{\centering\arraybackslash}p{0.2cm}>{\centering\arraybackslash}p{1.2cm}>{\centering\arraybackslash}p{1cm}>{\centering\arraybackslash}p{2.6cm}>{\centering\arraybackslash}p{1.4cm}}
1           & 6 &  5                             & 13 & 7                       \\ \hline
\multicolumn{1}{|@{}c@{}|}{0} & \multicolumn{1}{@{}c@{}|}{\textcolor{opcodecolor}{opcode}} & \multicolumn{1}{@{}c@{}|}{Sd} & \notableentry & \multicolumn{1}{@{}c@{}|}{Imm} \\ \hline
 & \tablescript{SMIS}  & \multicolumn{1}{@{}c@{}}{\tablescript{Dst SReg}} &  & \multicolumn{1}{@{}c@{}}{\tablescript{Qubit Mask}}
\end{tabular}
\begin{tabular}{>{\centering\arraybackslash}p{0.2cm}>{\centering\arraybackslash}p{1.2cm}>{\centering\arraybackslash}p{1cm}>{\centering\arraybackslash}p{0.8cm}>{\centering\arraybackslash}p{3.2cm}}
1           & 6 &  5                             & 4 & 16                       \\ \hline
\multicolumn{1}{|@{}c@{}|}{0} & \multicolumn{1}{@{}c@{}|}{\textcolor{opcodecolor}{opcode}} & \multicolumn{1}{@{}c@{}|}{Td} & \notableentry & \multicolumn{1}{@{}c@{}|}{Imm} \\ \hline
 & \tablescript{SMIT}  & \multicolumn{1}{@{}c@{}}{\tablescript{Dst TReg}} &  & \tablescript{Qubit Pair Mask}
\end{tabular}
\begin{tabular}{>{\centering\arraybackslash}p{0.2cm}>{\centering\arraybackslash}p{1.2cm}>{\centering\arraybackslash}p{1cm}>{\centering\arraybackslash}p{4.4cm}}
1           & 6 &  5                             & 20                       \\ \hline
\multicolumn{1}{|@{}c@{}|}{0} & \multicolumn{1}{@{}c@{}|}{\textcolor{opcodecolor}{opcode}} & \notableentry & \multicolumn{1}{@{}c@{}|}{Imm} \\ \hline
 & \tablescript{QWAIT}  &  & \tablescript{Wait time}
\end{tabular}
\begin{tabular}{>{\centering\arraybackslash}p{0.2cm}>{\centering\arraybackslash}p{1.2cm}>{\centering\arraybackslash}p{1cm}>{\centering\arraybackslash}p{1cm}>{\centering\arraybackslash}p{3cm}}
1           & 6 &  5                             & 5 & 15                       \\ \hline
\multicolumn{1}{|@{}c@{}|}{0} & \multicolumn{1}{@{}c@{}|}{\textcolor{opcodecolor}{opcode}} & \notableentry & \multicolumn{1}{@{}c@{}|}{Rs} & \notableentry \\ \hline
 & \tablescript{QWAITR}  &  & \multicolumn{1}{@{}c@{}}{\tablescript{Src GPR}} &
\end{tabular}
\begin{tabular}{>{\centering\arraybackslash}p{0.2cm}>{\centering\arraybackslash}p{1.8cm}>{\centering\arraybackslash}p{1cm}>{\centering\arraybackslash}p{1.8cm}>{\centering\arraybackslash}p{1cm}>{\centering\arraybackslash}p{0.2cm}}
1 & 9  & 5 & 9 & 5  & 3                       \\ \hline
\multicolumn{1}{|@{}c@{}|}{1} & \multicolumn{1}{@{}c@{}|}{\textcolor{opcodecolor}{q\_opcode}} & \multicolumn{1}{@{}c@{}|}{Si/Ti} & \multicolumn{1}{@{}c@{}|}{\textcolor{opcodecolor}{q\_opcode}} & \multicolumn{1}{@{}c@{}|}{Si/Ti} & \multicolumn{1}{@{}c@{}|}{PI} \\ \hline
 & \multicolumn{2}{@{}c@{}}{\tablescript{quantum operation 0}} & \multicolumn{2}{@{}c@{}}{\tablescript{quantum operation 1}} &
\end{tabular}
\caption{Format of the \lstinline!SMIS! and \lstinline!SMIT! (top two), \lstinline!QWAIT! and \lstinline!QWAITR! (middle two), and quantum bundle (bottom) instruction.}
\label{fig:bundle_fmt}
\end{figure}

\label{sec:uarch}
\begin{figure*}[bt]
\centering
\includegraphics[width=\textwidth]{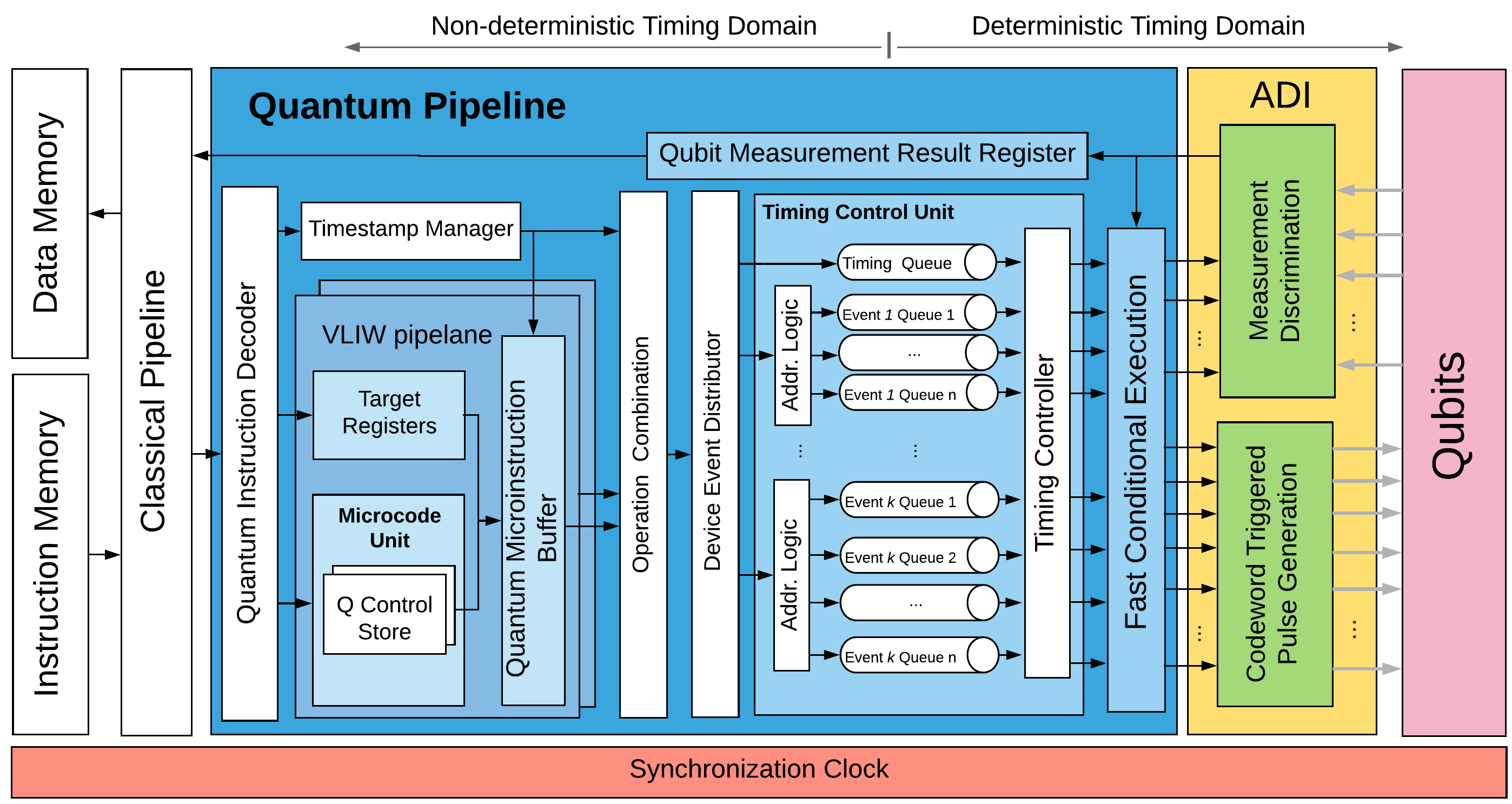}
\caption{Quantum microarchitecture implementing the instantiated eQASM for the seven-qubit superconducting quantum processor.}
\label{fig:qumav2}
\end{figure*}

There are 32 single- (two-)qubit target registers, and the target register address width is \bits{5}.
The target registers use a mask format.
The mask is 7- (16-)bit wide in the single- (two-)qubit target register.
Each bit in the mask of the value \bin{1} indicates that the corresponding qubit (allowed qubit pair) is selected.
In the \mbox{\lstinline!QWAIT(R)!} instruction, only the least significant \bits{20} of the \code{Imm} field or GPR \code{Rs} are used to specify the waiting time.
In the quantum bundle instruction, each quantum operation occupies \bits{14} and the q\_opcode is \bits{9}.

\subsection{Microarchitecture}
QuMIS is implemented by the control microarchitecture QuMA with codeword-based event control, queue-based event timing control and multi-level instruction decoding~\cite{fu2017experimental}.
Adopting these three mechanisms, we redesign a quantum control microarchitecture, QuMA\_v2, implementing the instantiated eQASM as shown in Fig.~\ref{fig:qumav2}.
It supports all features of eQASM.
The classical pipeline maintains the PC and implements the GPR file and the comparison flags.
The execution flag register is maintained by the fast conditional execution module.
The classical pipeline fetches and processes instructions one by one from the instruction memory.
All auxiliary classical instructions are processed by the classical pipeline while quantum instructions are forwarded to the quantum pipeline for further processing.

The timestamp manager processes the \lstinline!QWAIT(R)! instructions and the \code{PI} field to generate timing points.
The quantum pipeline contains a VLIW front end with two VLIW lanes, each lane processing one quantum operation.
The \lstinline!SMIS! (\lstinline!SMIT!) instructions update the corresponding target registers in each VLIW lane.
Inside each VLIW lane, the q\_opcode is translated by the microcode unit into one micro-operation (labeled as $\mu\_op_{\mathrm{s}}$) for a single-qubit operation or two micro-operations (labeled as $\mu\_op_{\mathrm{src}}$ and $\mu\_op_{\mathrm{tgt}}$) for a two-qubit operation.
$\mu\_op_{\mathrm{src}}$ ($\mu\_op_{\mathrm{tgt}}$) will be applied on the source (target) qubit of the target qubit pair.
The configuration of the microcode unit is stored in the Q control store, which is implemented using a lookup table.
The target register \code{Si} (\code{Ti}) is read for a single- (two-)qubit operation.

The quantum microinstruction buffer resolves the mask-based qubit address and associates the quantum operations to the last generated timing point.
It resolves the qubit address in two steps.

First, the mask stored in \code{Si} (\code{Ti}) is translated into seven two-bit micro-operation selection signals $\mathrm{OpSel}_i$, where $i=0,1,\cdots,6$, with each signal for one qubit.
Table~\ref{tab:opsel} lists the meaning of every case of the micro-operation selection signal.
\begin{table}[hbt]
    \centering
    \small
    \caption{Definition of the micro-operation selection signal.}
    \label{tab:opsel}
    \begin{tabular}{|c|c|c|c|}
    \hline
        Value & Operation to Select & Value & Operation to Select \\ \hline
        \bin{00} & None & \bin{10} & $\mu\_op_{\mathrm{tgt}}$\\ \hline
        \bin{01} & $\mu\_op_{\mathrm{src}}$ & \bin{11} & $\mu\_op_{\mathrm{s}}$\\ \hline
    \end{tabular}
\end{table}
For single-qubit operations, $\mathrm{OpSel}_i$ is set to \bin{11} (\bin{00}) if the $i$-th bit in the mask is \bin{1} (\bin{0}).
For a two-qubit operation, $\mathrm{OpSel}_i$ is set to \bin{00} if qubit $i$ is not contained in any selected allowed qubit pair.
Otherwise, $\mathrm{OpSel}_i$ is \bin{01} (\bin{10}) if the target qubit pair contains qubit $i$ as the source (target) qubit.
Take qubit 0 as an example.
It is connected to edges 0, 1, 8, and 9.
When edge 0 or 9 (1 or 8) is selected in the mask, qubit 0 is the target (source) qubit and should be applied with $\mu\_op_{\mathrm{tgt}}$ ($\mu\_op_{\mathrm{src}}$).
In other words, $\mathrm{OpSel}_0$ should be \bin{10} (\bin{01}), and can be generated using a simple \code{OR} ($\vee$) logic:
\begin{center}
    $\mathrm{OpSel}_0$ = (\code{Ti}[0]~$\vee$~\code{Ti}[9])~::~(\code{Ti}[1]~$\vee$~\code{Ti}[8]).
\end{center}
The assembler should check the validity of two-qubit target register values.
For example, it is invalid if two edges connecting to the same qubit are selected in the same \code{T} register.

Second, based on $\mathrm{OpSel}_i$, either none or one micro-operation is output for qubit $i$.
This step is fully parallel.

The operation combination module also works in a two-step fashion.
First, since each VLIW lane outputs none or one micro-operation for each qubit, the operation combination module merges both micro-operations from both VLIW lanes.
If both VLIW lanes output one micro-operation on the same qubit, an error is raised, and the quantum processor stops.
Second, as explained in Section~\ref{sec:vliw}, a long quantum bundle requires multiple quantum bundle instructions to describe it.
The operation combination module buffers all micro-operations associated with the same timing point.
Only when it detects that all quantum operations in the same quantum bundle have been collected, the operation combination module sends the buffered micro-operations to the device event distributor.
This detection can be done, e.g., by recognizing a new timing point generated by the timestamp manager which is different to the one associated to the buffered micro-operations.
Also, if two different quantum bundle instructions specify a quantum operation on the same qubit, an error is raised, and the quantum processor stops.

As shown in Section~\ref{sec:implementation}, operating a qubit may require the collaboration of multiple electronic devices in the analog-digital-interface, and a single device may also control multiple qubits.
Hence, the micro-operations should be reorganized into \textit{device operations} to trigger the corresponding devices.
The device event distributor reorganizes multiple micro-operations associated with the same timing label into different device operations.
After that, each device operation with the associated timing label is buffered at an event queue of the timing control unit awaiting execution.
The timing controller then triggers every device operation at its expected timing point.

After the device operations have been triggered by the timing controller, fast conditional execution is performed based on the selected execution flags of the target qubits.
The execution flag selection signal comes from the microcode unit configured by the programmer.
Only device operations for qubits of which the selected execution flag is \bin{1} are released to the analog-digital interface (ADI).
In this eQASM instantiation, four types of combinatorial logic are used to define the execution flags:
\begin{enumerate}
    \item \bin{1} (the default for unconditional execution);
    \item \bin{1} \textbf{iff} the last finished measurement result is $\ket{1}$;
    \item \bin{1} \textbf{iff} the last finished measurement result is $\ket{0}$;
    \item \bin{1} \textbf{iff} the last two finished measurements get the same result.
\end{enumerate}
Note, the last finished measurement result refers to the result of the last finished measurement instruction on this qubit when these flags are used.
It is irrelevant to the validity of the quantum measurement result register.
Once there returns a measurement result for a qubit from the analog-digital interface, the fast conditional execution unit immediately update the execution flags corresponding to that qubit.

To support CFC, a counter \code{Ci} is attached to each qubit measurement result register \code{Qi}, with an initial value of 0.
Once a measurement instruction acting on qubit $i$ is issued from the classical pipeline to the quantum pipeline, \code{Ci} increments by 1.
If the measurement discrimination unit writes back a measurement result for qubit $i$, \code{Ci} decrements by 1.
\code{Qi} is valid only when \code{Ci} is 0.
If \code{Ci} is not 0 when the instruction \lstinline[basicstyle=\small\ttfamily]!FMR Rd, Qi! is issued, the pipeline is stalled until \code{Ci} is 0.
In this way, it is ensured that the instruction
\mbox{\lstinline[basicstyle=\small\ttfamily]!FMR Rd, !}\allowbreak\lstinline[basicstyle=\small\ttfamily]!Qi!  
always fetches the result of the last measurement instruction acting on qubit $i$.

\subsection{Implementation}
\label{sec:implementation}
\begin{figure}[bt]
\centering
\includegraphics[width=\columnwidth]{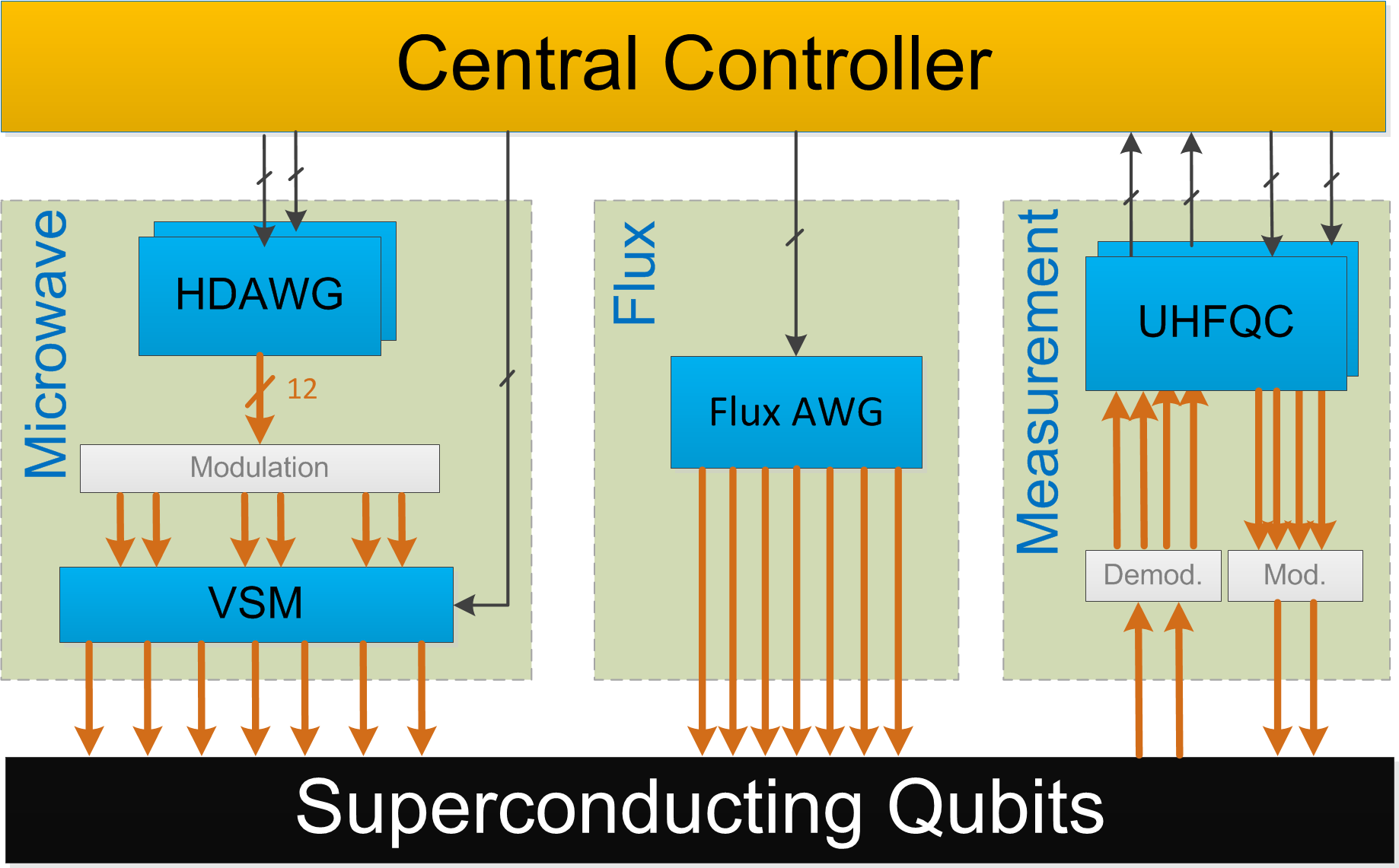}
\caption{Hardware structure implementing the instantiated eQASM for the seven-qubit superconducting quantum processor.
Thin (thick) lines represent digital (analog) signals.}
\label{fig:qusurf}
\end{figure}

The hardware structure implementing the microarchitecture (Fig.~\ref{fig:qusurf}) consists of a Central Controller responsible for orchestrating three modules containing slave devices for microwave control, flux control, and measurement.

The Central Controller is a digital device built with an Intel Altera Cyclone V SOC 5CSTFD6D5F31I7N Field Programmable Gate Array (FPGA) chip.
The Central Controller implements the digital part of the microarchitecture (left to the ADI in Fig.~\ref{fig:qumav2}).
The timing controller and fast conditional execution module work at \MHz{50} to get a cycle time of \ns{20}.
The other parts work at \MHz{100}.

Single-qubit $x$ and $y$ rotations are performed by applying microwave pulses to the qubits.
The pulses are generated by Zurich Instruments High Density Arbitrary Waveform Generators (HDAWG) and modulated using a Rohde \& Schwarz (R\&S) SGS100A microwave source.
A custom-built vector switch matrix (VSM) is responsible for duplicating and routing the pulses to the respective qubits as well as tailoring the waveforms to the individual qubits~\cite{asaad2016independent} using a qubit-frequency reuse scheme that allows for efficient scaling of the microwave control module~\cite{versluis2017scalable}.

Flux pulses that implement two-qubit CZ gates and single-qubit $z$ rotations are performed by applying pulses generated by an HDAWG on the dedicated flux lines for each qubit.

The measurement discrimination unit is implemented using two Zurich Instruments Ultra-High-Frequency Quantum Controllers (UHFQC) connected to the two feedlines shown in Fig.
~\ref{fig:topo}.
The UHFQC has two analog outputs that can be used to generate the measurement pulses and two analog inputs to sample the transmitted signals from which the UHFQC can infer the measurement result.
The measurement pulses going to (coming from) the qubits are modulated (demodulated) using a single R\&S SGS100A.
All analog ports operate at \GSps{1.8} allowing for simultaneous measurement of up to 9 qubits per feedline using frequency multiplexing techniques ~\cite{heinsoo2018rapid}.

The Central Controller connects to the UHFQCs and HDAWGs via a 32-bit digital interface working at \MHz{50}.
Since measurement results are sent from the UHFQC to the Central Controller, \bits{16} of the connection are sent from the Central Controller to the UHFQC and the other \bits{16} the other way around.
All operations on UHFQCs and HDAWGs are codeword triggered.
The routing of microwave pulses by the VSM is controlled through seven digital signals with a sampling rate of 400 MSa/s.

\section{Experiment}
\label{sec:experiment}
Since the target seven-qubit quantum chip is still under test at the time of writing, we replaced the quantum chip of this microarchitecture with a two-qubit superconducting quantum processor to validate the eQASM design.
The two qubits are interconnected and coupled to a single feedline.
A configuration file is used to specify the quantum chip topology with the two qubits renamed as qubit 0 and 2.
It is used by the quantum compiler and the assembler.
eQASM programs used to perform the experiments as described below are all compiled from OpenQL descriptions with corresponding quantum operation configuration.

\begin{figure}[bt]
    \centering
    \includegraphics[width=\columnwidth]{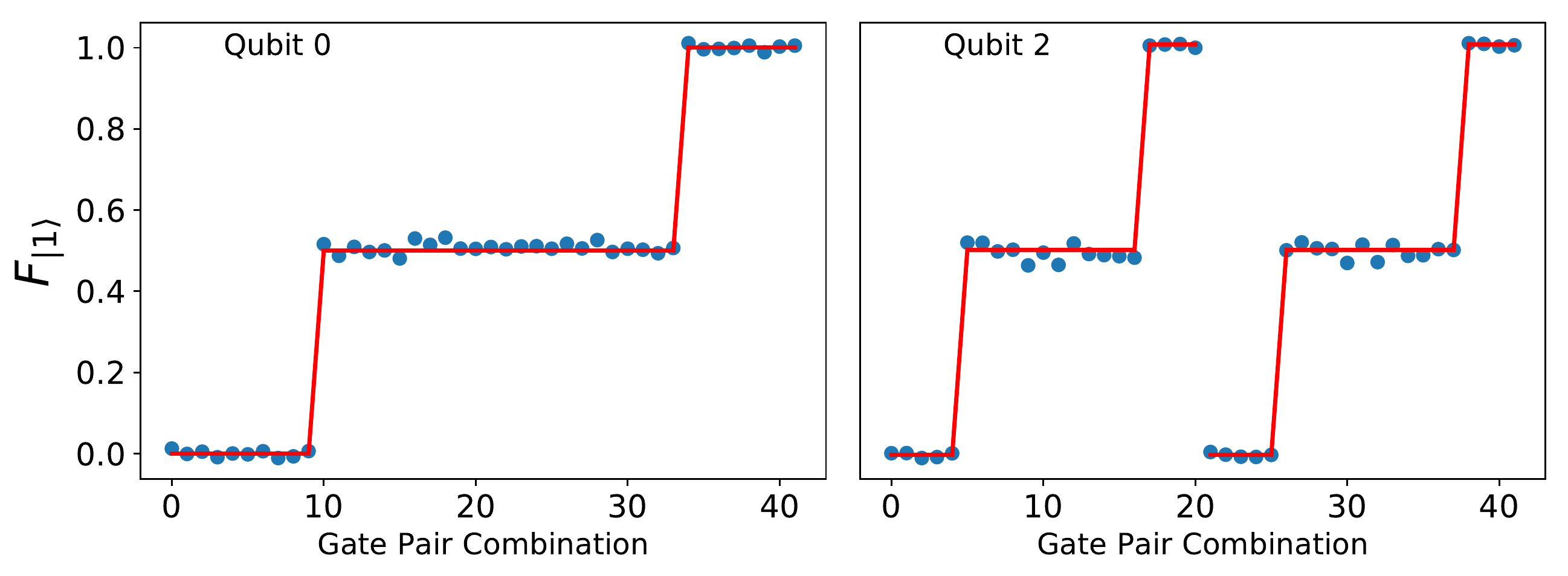}
    \caption{Two-qubit \textit{AllXY} result, corrected for readout errors.}
    \label{fig:twoqubitallxyresult}
\end{figure}

We first used eQASM to perform some single-qubit calibration experiments which utilize uncalibrated operations.
For example, the Rabi oscillation~\cite{reed2013entanglement} applies an $x$-rotation pulse on the qubit after initialization and then measures it.
A sequence of fixed-length $x$-rotation pulses with variable amplitudes are used.
Each pulse in the sequence is uploaded to the codeword triggered pulse generation unit of the microarchitecture and configured to be an operation \mbox{\lstinline!X_Amp_i!} in eQASM.
As a result, this experiment calibrated the amplitude of the $X$ gate pulse.
Together with other experiments, the fidelity of single-qubit quantum operations used later reached 99.90\% as measured in the following RB experiment.
It is worth mentioning that we observed considerable speedup in performing these experiments with the eQASM control paradigm in practice.

eQASM is then configured to include single-qubit gates $\{I,\allowbreak X,\allowbreak Y,\allowbreak X_{90},\allowbreak Y_{90},\allowbreak X_{\mathrm{m}90},\allowbreak Y_{\mathrm{m}90}\}$ and a two-qubit CZ gate for the following experiments.
The \textit{AllXY} experiment is typically used to calibrate single-qubit gates.
In \textit{AllXY}, pairs of single-qubit gates are chosen from the set $\{I,\allowbreak X,\allowbreak Y,\allowbreak X_{90},\allowbreak Y_{90}\}$ and applied in such a way that the expected measurement outcomes produce a characteristic staircase pattern that is highly sensitive to gate errors (red line in Fig.~\ref{fig:twoqubitallxy}).
In the two-qubit \textit{AllXY} experiment, the control pulses are applied on each qubit simultaneously.
The sequence is modified to distinguish the qubits on which it is applied: each gate pair in the sequence is repeated on the first qubit while the entire sequence is repeated on the second qubit.
The fidelity of qubit to the $\ket{1}$ state can be extracted by averaging the measurement results for each gate pair over $N$ rounds and correcting for readout errors.
The eQASM program for one routine of this experiment is shown in Fig.~\ref{fig:twoqubitallxy}.
Figure~\ref{fig:twoqubitallxyresult} shows the final measurement result of the entire experiment (blue dots), which matches well with the expectation (red line).
This demonstrates that the timing control, SOMQ, and VLIW of eQASM work properly in the experiment.

To evaluate the impact of the timing of operations on the error rate, we use single-qubit randomized benchmarking, a technique that can estimate the average error rate for a set of operations under a very general noise model~\cite{magesan2011scalable, epstein2014investigating}.
In this experiment, a sequence of $k$ random Clifford gates are applied on a qubit initialized in the $\ket{0}$ state.
Before measurement, a Clifford is chosen that inverts all preceding operations so that the qubit should end up in the $\ket{0}$ state with survival probability $p(k)$.
By performing this experiment for different $k$ and averaging over many randomizations, the Clifford fidelity $F_{\mathrm{Cl}}$ can be extracted from the exponential decay.
Because each Clifford gate is decomposed into primitive $x$- and $y$-rotations the gate count is increased by 1.875 on average.
The average error rate per gate, $\epsilon$, is then calculated as $\epsilon=1-F_{\mathrm{Cl}}^{1/1.875}$.

\begin{figure}[bt]
    \centering
    \includegraphics[width=\columnwidth]{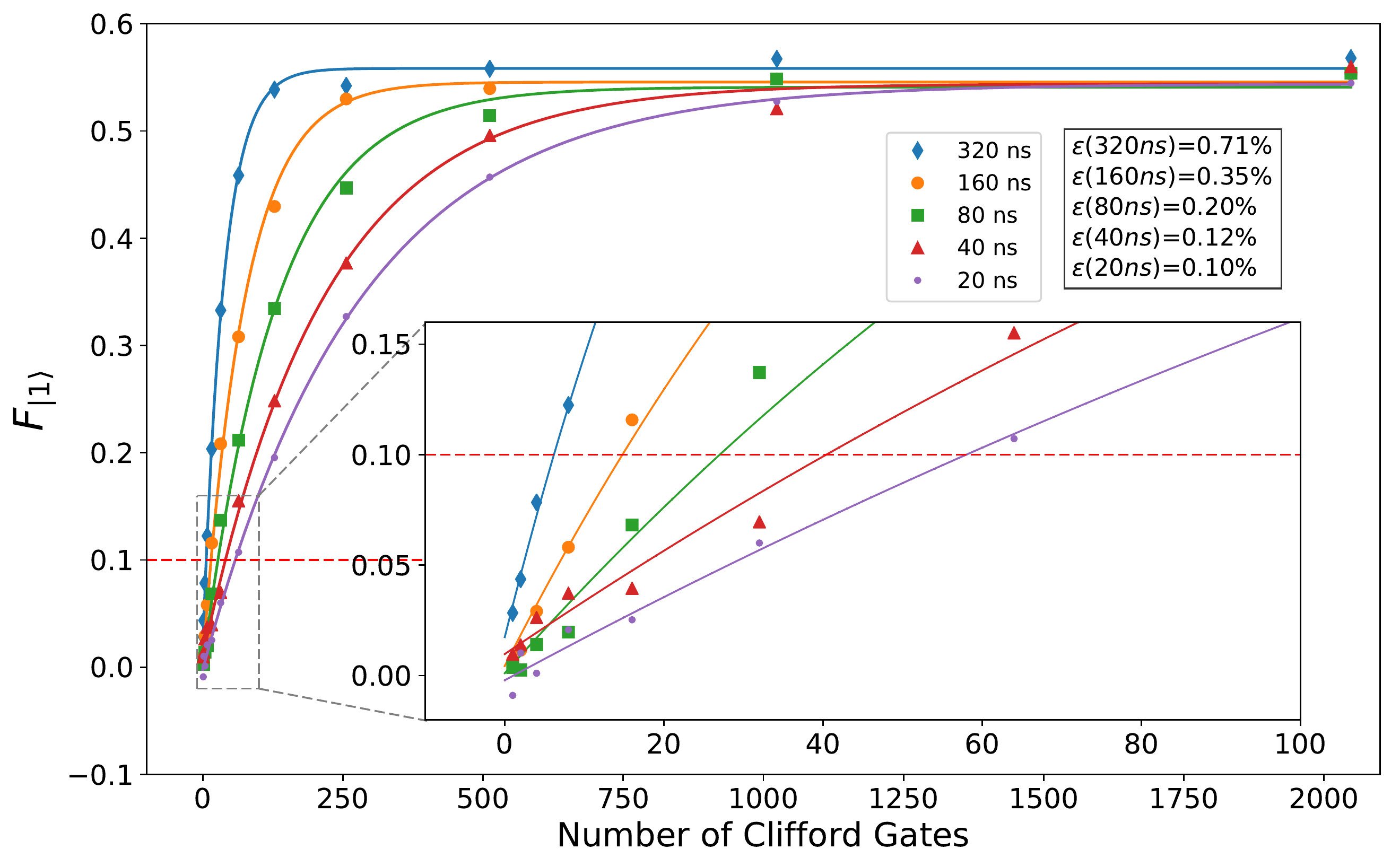}
    \caption{Single-qubit randomized benchmarking results for different intervals between gates.
    Dashed line indicates a 10\% error rate for visual reference.
    }
    \label{fig:rb}
\end{figure}

Single-qubit randomized benchmarking was performed for different intervals between the starting points of consecutive gates (320, 160, 80, 40, and \ns{20}).
As shown in Fig.~\ref{fig:rb}, the average error per gate decreases by a factor of $\sim7$, from 0.71\% to 0.10\% when decreasing the interval from \ns{320} to \ns{20}.
This demonstrates the significant impact of timing on the fidelity of the final computation result, which substantiates the requirement of explicit specification of timing at QISA level to enable platform-specific optimization and especially scheduling by the compiler.

Fast conditional execution is verified by the active qubit reset experiment with qubit 2 using the code as shown in Fig.~\ref{fig:ffc}.
We find the probability of measuring the qubit in the $\ket{0}$ state after conditionally applying the \lstinline!C_X! gate to be 82.7\%, limited by the readout fidelity.
We verified CFC by connecting the Central Controller and the UHFQC.
The eQASM program used is shown in Fig.~\ref{fig:cfc}.
The UHFQC is programmed to generate alternative mock measurement results for qubit 0.
The alternation between $X$ and $Y$ operations is verified by detecting the output digital signals using an oscilloscope.
We also measured the feedback latency of fast conditional execution and CFC, which are \mbox{\ns{\sim92} and \ns{\sim316}}, respectively.
The feedback latency is defined as the time between sending the measurement result into the Central Controller and receiving the digital output based on the feedback from the Central Controller.

As a proof of concept of performing quantum algorithms using eQASM, we executed a two-qubit Grover's search algorithm~\cite{grover1996fast, dicarlo2009demonstration}.
The algorithmic fidelity, i.e., correcting for readout infidelity, is found to be 85.6\% using quantum tomography with maximum likelihood estimation.
This fidelity is limited by the CZ gate.

\section{Conclusion}
\label{sec:conclusion}
In this paper, we have proposed eQASM, a QISA that can be directly executed on a control microarchitecture after instantiation.
With runtime feedback, eQASM supports full quantum program flow control at the (micro)architecture level~\cite{selinger2004towards, ying2016foundations}.
With efficient timing specification, SOMQ execution, and VLIW architecture, eQASM alleviates the quantum operation issue rate problem, presenting better scalability than QuMIS.
Quantum operations in eQASM can be configured at compile time instead of QISA design time, which can support uncalibrated or uncommon operations, leaving ample space for compiler-based optimization.
Low-level hardware information mainly appears in the binary of a particular eQASM instantiation, which makes eQASM assembly expressive.
It is worth noting that by removing the timing information in the eQASM description, the quantum semantics of the program can be kept and further converted into another executable format targeting another hardware platform.

As validation, eQASM was instantiated into a 32-bit instruction set targeting a seven-qubit superconducting quantum chip, and implemented using a quantum microarchitecture.
eQASM was verified by several experiments with this microarchitecture performed on a two-qubit chip.
The efficiency improvement observed in using eQASM to control quantum experiments broadens the scope of application of quantum assemblies.

Future work will include performing verifying comprehensive feedback control with qubits and controlling the originally targeted seven-qubit superconducting quantum processor with the implemented microarchitecture.
Also, it will be interesting to instantiate eQASM to control other quantum processors, including superconducting quantum processors with a different quantum chip topology, and altogether different quantum hardware, such as spins in quantum dots~\cite{watson2018programmable}, nitrogen vacancy centers~\cite{cramer2016repeated}.

\balance
\section*{Acknowledgements}
We thank C.~C.~Bultink, L.~Lao, and M.~Zhang for discussions while designing and implementing eQASM, S.~Poletto, N.~Haider, D.~J.~Michalak, and A.~Bruno for designing and fabricating the two-qubit quantum chip used in experiments, T.~van~der~Laan, and J.~Veltin for project management when implementing the Central Controller, Y.~Shi for discussions on feedback control. We acknowledge funding from the China Scholarship Council (X.~Fu), Intel Corporation, an ERC Synergy Grant, and the Office of the Director of National Intelligence (ODNI), Intelligence Advanced Research Projects Activity (IARPA), via the U.S. Army Research Office grant W911NF-16-1-0071. The views and conclusions contained herein are those of the authors and should not be interpreted as necessarily representing the official policies or endorsements, either expressed or implied, of the ODNI, IARPA, or the U.S. Government. The U.S. Government is authorized to reproduce and distribute reprints for Governmental purposes notwithstanding any copyright annotation thereon.

\bibliographystyle{IEEEtran.bst}
\bibliography{reference}

\begin{thebibliography}{10}
\providecommand{\url}[1]{#1}
\csname url@samestyle\endcsname
\providecommand{\newblock}{\relax}
\providecommand{\bibinfo}[2]{#2}
\providecommand{\BIBentrySTDinterwordspacing}{\spaceskip=0pt\relax}
\providecommand{\BIBentryALTinterwordstretchfactor}{4}
\providecommand{\BIBentryALTinterwordspacing}{\spaceskip=\fontdimen2\font plus
\BIBentryALTinterwordstretchfactor\fontdimen3\font minus
  \fontdimen4\font\relax}
\providecommand{\BIBforeignlanguage}[2]{{%
\expandafter\ifx\csname l@#1\endcsname\relax
\typeout{** WARNING: IEEEtran.bst: No hyphenation pattern has been}%
\typeout{** loaded for the language `#1'. Using the pattern for}%
\typeout{** the default language instead.}%
\else
\language=\csname l@#1\endcsname
\fi
#2}}
\providecommand{\BIBdecl}{\relax}
\BIBdecl

\bibitem{fu2017experimental}
X.~Fu, M.~A. Rol, C.~C. Bultink, J.~van Someren, N.~Khammassi, I.~Ashraf,
  R.~F.~L. Vermeulen, J.~C. de~Sterke, W.~J. Vlothuizen, R.~N. Schouten, C.~G.
  Almudever, L.~DiCarlo, and K.~Bertels, ``An experimental microarchitecture
  for a superconducting quantum processor,'' in \emph{Proceedings of the 50th
  Annual IEEE/ACM International Symposium on Microarchitecture}.\hskip 1em plus
  0.5em minus 0.4em\relax ACM, 2017, pp. 813--825.

\bibitem{feynman1982simulating}
R.~P. Feynman, ``Simulating physics with computers,'' \emph{International
  Journal of Theoretical Physics}, vol.~21, pp. 467--488, 1982.

\bibitem{kassal2011simulating}
I.~Kassal, J.~D. Whitfield, A.~Perdomo-Ortiz, M.-H. Yung, and A.~Aspuru-Guzik,
  ``Simulating chemistry using quantum computers,'' \emph{Annual Review of
  Physical Chemistry}, vol.~62, pp. 185--207, 2011.

\bibitem{preskill2018quantum}
J.~Preskill, ``Quantum computing in the {NISQ} era and beyond,''
  \emph{arXiv:1801.00862}, 2018.

\bibitem{terhal2015quantum}
B.~M. Terhal, ``Quantum error correction for quantum memories,'' \emph{Reviews
  of Modern Physics}, vol.~87, p. 307, 2015.

\bibitem{selinger2004towards}
P.~Selinger, ``Towards a quantum programming language,'' \emph{Mathematical
  Structures in Computer Science}, vol.~14, pp. 527--586, 2004.

\bibitem{riste2012feedback}
D.~Rist{\`e}, C.~C. Bultink, K.~W. Lehnert, and L.~DiCarlo, ``Feedback control
  of a solid-state qubit using high-fidelity projective measurement,''
  \emph{Physical Review Letters}, vol. 109, p. 240502, 2012.

\bibitem{bennett1993teleporting}
C.~H. Bennett, G.~Brassard, C.~Cr{\'e}peau, R.~Jozsa, A.~Peres, and W.~K.
  Wootters, ``Teleporting an unknown quantum state via dual classical and
  einstein-podolsky-rosen channels,'' \emph{Physical Review Letters}, vol.~70,
  p. 1895, 1993.

\bibitem{paetznick2014repeat}
A.~Paetznick and K.~M. Svore, ``{Repeat-Until-Success: Non-deterministic
  decomposition of single-qubit unitaries},'' \emph{Quantum Information \&
  Computation}, vol.~{14}, pp. 1277--1301, {2014}.

\bibitem{kitaev1996quantum}
A.~Y. Kitaev, ``Quantum measurements and the {Abelian} stabilizer problem,'' in
  \emph{Electronic Colloquium on Computational Complexity}, 1996.

\bibitem{javadiabhari2015scaffcc}
A.~JavadiAbhari, S.~Patil, D.~Kudrow, J.~Heckey, A.~Lvov, F.~T. Chong, and
  M.~Martonosi, ``{ScaffCC: Scalable compilation and analysis of quantum
  programs},'' \emph{Parallel Computing}, vol.~45, pp. 2--17, 2015.

\bibitem{smith2016practical}
R.~S. Smith, M.~J. Curtis, and W.~J. Zeng, ``A practical quantum instruction
  set architecture,'' \emph{arXiv:1608.03355}, 2016.

\bibitem{liu2017qsi}
S.~Liu, X.~Wang, L.~Zhou, J.~Guan, Y.~Li, Y.~He, R.~Duan, and M.~Ying,
  ``{Q$|$SI$\rangle$: A quantum programming environment},''
  \emph{arXiv:1710.09500}, 2017.

\bibitem{cross2017open}
A.~W. Cross, L.~S. Bishop, J.~A. Smolin, and J.~M. Gambetta, ``Open quantum
  assembly language,'' \emph{arXiv:1707.03429}, 2017.

\bibitem{chong2017programming}
F.~T. Chong, D.~Franklin, and M.~Martonosi, ``Programming languages and
  compiler design for realistic quantum hardware,'' \emph{Nature}, vol. 549, p.
  180, 2017.

\bibitem{ying2016foundations}
M.~Ying, \emph{Foundations of Quantum Programming}.\hskip 1em plus 0.5em minus
  0.4em\relax Morgan Kaufmann, 2016.

\bibitem{wecker2014liqui}
D.~Wecker and K.~M. Svore, ``{LIQUi$\mid\rangle$: A software design
  architecture and domain-specific language for quantum computing},''
  \emph{arXiv:1402.4467}, 2014.

\bibitem{Steiger2016projectq}
D.~S. Steiger, T.~H{\"a}ner, and M.~Troyer, ``{ProjectQ: an open source
  software framework for quantum computing},'' \emph{Quantum}, vol.~2, 2018.

\bibitem{ibmq}
{IBM Q team}, ``{IBM Quantum Experience},''
  \url{https://www.research.ibm.com/ibm-q/}, 2018.

\bibitem{rigettiforest}
{Rigetti}, ``{Rigetti Forest},'' \url{https://www.rigetti.com/forest}, 2018.

\bibitem{alibabaquantumcloud}
{Alibaba}, ``{Alibaba Quantum Computing Cloud},''
  \url{http://quantumcomputer.ac.cn}, 2018.

\bibitem{bultink2016active}
C.~C. Bultink, M.~A. Rol, T.~E. O'Brien, X.~Fu, B.~Dikken, R.~Vermeulen, J.~C.
  de~Sterke, A.~Bruno, R.~N. Schouten, and L.~DiCarlo, ``{Active resonator
  reset in the nonlinear dispersive regime of circuit QED},'' \emph{Physical
  Review Applied}, vol.~6, p. 034008, 2016.

\bibitem{ofek2016extending}
N.~Ofek, A.~Petrenko, R.~Heeres, P.~Reinhold, Z.~Leghtas, B.~Vlastakis, Y.~Liu,
  L.~Frunzio, S.~Girvin, L.~Jiang \emph{et~al.}, ``Extending the lifetime of a
  quantum bit with error correction in superconducting circuits,''
  \emph{Nature}, vol. 536, p. 441, 2016.

\bibitem{hu2018demonstration}
L.~Hu, Y.~Ma, W.~Cai, X.~Mu, Y.~Xu, W.~Wang, Y.~Wu, H.~Wang, Y.~Song, C.~Zou,
  S.~M. Girvin, L.-M. Duan, and L.~Sun, ``Demonstration of quantum error
  correction and universal gate set on a binomial bosonic logical qubit,''
  \emph{arXiv:1805.09072}, 2018.

\bibitem{fu2018microarchitecture}
X.~Fu, M.~A. Rol, C.~C. Bultink, J.~van Someren, N.~Khammassi, I.~Ashraf,
  R.~F.~L. Vermeulen, J.~C. de~Sterke, W.~J. Vlothuizen, R.~N. Schouten, C.~G.
  Almudever, L.~DiCarlo, and K.~Bertels, ``A microarchitecture for a
  superconducting quantum processor,'' \emph{IEEE Micro}, vol.~38, pp. 40--47,
  2018.

\bibitem{tannu2017taming}
S.~S. Tannu, Z.~A. Myers, P.~J. Nair, D.~M. Carmean, and M.~K. Qureshi,
  ``Taming the instruction bandwidth of quantum computers via hardware-managed
  error correction,'' in \emph{Proceedings of the 50th Annual IEEE/ACM
  International Symposium on Microarchitecture}.\hskip 1em plus 0.5em minus
  0.4em\relax IEEE/ACM, 2017, pp. 679--691.

\bibitem{ryan2017hardware}
C.~A. Ryan, B.~R. Johnson, D.~Rist{\`e}, B.~Donovan, and T.~A. Ohki, ``Hardware
  for dynamic quantum computing,'' \emph{arXiv:1704.08314}, 2017.

\bibitem{nkhammassi2018cqasm}
N.~Khammassi, G.~G. Guerreschi, I.~Ashraf, J.~W. Hogaboam, C.~G. Almudever, and
  K.~Bertels, ``{cQASM} v1.0: Towards a common quantum assembly language,''
  \emph{arXiv:1805.09607}, 2018.

\bibitem{stone2010opencl}
J.~E. Stone, D.~Gohara, and G.~Shi, ``{OpenCL: A parallel programming standard
  for heterogeneous computing systems},'' \emph{Computing in Science \&
  Engineering}, vol.~12, pp. 66--73, 2010.

\bibitem{abhari2012scaffold}
A.~J. Abhari, A.~Faruque, M.~J. Dousti, L.~Svec, O.~Catu, A.~Chakrabati, C.-F.
  Chiang, S.~Vanderwilt, J.~Black, and F.~Chong, ``Scaffold: Quantum
  programming language,'' Princeton University, Technical Report, 2012.

\bibitem{svore2018q}
K.~Svore, A.~Geller, M.~Troyer, J.~Azariah, C.~Granade, B.~Heim,
  V.~Kliuchnikov, M.~Mykhailova, A.~Paz, and M.~Roetteler, ``Q\#: Enabling
  scalable quantum computing and development with a high-level {DSL},'' in
  \emph{Proceedings of the Real World Domain Specific Languages Workshop
  2018}.\hskip 1em plus 0.5em minus 0.4em\relax ACM, 2018, p.~7.

\bibitem{mckay2018qiskit}
D.~C. McKay, T.~Alexander, L.~Bello, M.~J. Biercuk, L.~Bishop, J.~Chen, J.~M.
  Chow, A.~D. C{\'o}rcoles, D.~Egger, S.~Filipp, J.~Gomez, M.~Hush,
  A.~Javadi-Abhari, D.~Moreda, P.~Nation, B.~Paulovicks, E.~Winston, C.~J.
  Wood, J.~Wootton, and J.~M. Gambetta, ``Qiskit backend specifications for
  {OpenQASM} and {OpenPulse} experiments,'' \emph{arXiv:1809.03452}, 2018.

\bibitem{werschnik2007quantum}
J.~Werschnik and E.~Gross, ``Quantum optimal control theory,'' \emph{Journal of
  Physics B: Atomic, Molecular and Optical Physics}, vol.~40, p. R175, 2007.

\bibitem{leung17speedup}
N.~Leung, M.~Abdelhafez, J.~Koch, and D.~Schuster, ``Speedup for quantum
  optimal control from automatic differentiation based on graphics processing
  units,'' \emph{Physical Review A}, vol.~95, p. 042318, 2017.

\bibitem{lee2009computing}
E.~A. Lee, ``Computing needs time,'' \emph{Communications of the ACM}, vol.~52,
  pp. 70--79, 2009.

\bibitem{deutsch1995universality}
D.~Deutsch, A.~Barenco, and A.~Ekert, ``Universality in quantum computation,''
  \emph{Proceedings of the Royal Society of London A: Mathematical, Physical
  and Engineering Sciences}, vol. 449, pp. 669--677, 1995.

\bibitem{lloyd1995almost}
S.~Lloyd, ``Almost any quantum logic gate is universal,'' \emph{Physical Review
  Letters}, vol.~75, p. 346, 1995.

\bibitem{divincenzo1998quantum}
D.~P. DiVincenzo, ``Quantum gates and circuits,'' \emph{Proceedings of the
  Royal Society of London A: Mathematical, Physical and Engineering Sciences},
  vol. 454, pp. 261--276, 1998.

\bibitem{kudrow2013quantum}
D.~Kudrow, K.~Bier, Z.~Deng, D.~Franklin, Y.~Tomita, K.~R. Brown, and F.~T.
  Chong, ``Quantum rotations: a case study in static and dynamic machine-code
  generation for quantum computers,'' in \emph{ACM SIGARCH Computer
  Architecture News}.\hskip 1em plus 0.5em minus 0.4em\relax ACM, 2013, pp.
  166--176.

\bibitem{heckey2015compiler}
J.~Heckey, S.~Patil, A.~JavadiAbhari, A.~Holmes, D.~Kudrow, K.~R. Brown,
  D.~Franklin, F.~T. Chong, and M.~Martonosi, ``Compiler management of
  communication and parallelism for quantum computation,'' in \emph{Proceedings
  of 20th ACM International Conference on Architectural Support for Programming
  Languages and Operating Systems}.\hskip 1em plus 0.5em minus 0.4em\relax ACM,
  2015, pp. 445--456.

\bibitem{vassiliadis2003microcode}
S.~Vassiliadis, S.~Wong, and S.~Cotofana, ``Microcode processing: Positioning
  and directions,'' \emph{{IEEE Micro}}, vol.~{23}, pp. 21--30, {2003}.

\bibitem{flynn1972some}
M.~J. Flynn, ``Some computer organizations and their effectiveness,''
  \emph{IEEE transactions on computers}, vol. 100, pp. 948--960, 1972.

\bibitem{debnath2016demonstration}
S.~Debnath, N.~Linke, C.~Figgatt, K.~Landsman, K.~Wright, and C.~Monroe,
  ``Demonstration of a small programmable quantum computer with atomic
  qubits,'' \emph{Nature}, vol. 536, pp. 63--66, 2016.

\bibitem{ibmqx2}
{$5$-qubit backend: IBM Q team}, ``{IBM Q 5 Yorktown backend specification
  V1.1.0},'' Retrieved from
  \url{https://github.com/QISKit/qiskit-backend-information/tree/master/backends/yorktown/V1},
  2018.

\bibitem{balensiefer2005evaluation}
S.~Balensiefer, L.~Kregor-Stickles, and M.~Oskin, ``An evaluation framework and
  instruction set architecture for ion-trap based quantum
  micro-architectures,'' in \emph{Proceedings of 32nd International Symposium
  on Computer Architecture}.\hskip 1em plus 0.5em minus 0.4em\relax IEEE, 2005,
  pp. 186--196.

\bibitem{versluis2017scalable}
R.~Versluis, S.~Poletto, N.~Khammassi, B.~Tarasinski, N.~Haider, D.~J.
  Michalak, A.~Bruno, K.~Bertels, and L.~DiCarlo, ``Scalable quantum circuit
  and control for a superconducting surface code,'' \emph{Physical Review
  Applied}, vol.~8, p. 034021, 2017.

\bibitem{fowler2012surface}
A.~G. Fowler, M.~Mariantoni, J.~M. Martinis, and A.~N. Cleland, ``Surface
  codes: Towards practical large-scale quantum computation,'' \emph{Physical
  Review A}, vol.~86, p. 032324, 2012.

\bibitem{riesebos2017pauli}
L.~Riesebos, X.~Fu, S.~Varsamopoulos, C.~G. Almudever, and K.~Bertels, ``Pauli
  frames for quantum computer architectures,'' in \emph{Proceedings of the 54th
  Annual Design Automation Conference}.\hskip 1em plus 0.5em minus 0.4em\relax
  ACM, 2017, p.~76.

\bibitem{magesan2011scalable}
E.~Magesan, J.~M. Gambetta, and J.~Emerson, ``Scalable and robust randomized
  benchmarking of quantum processes,'' \emph{Physical Review Letters}, vol.
  106, p. 180504, 2011.

\bibitem{epstein2014investigating}
J.~M. Epstein, A.~W. Cross, E.~Magesan, and J.~M. Gambetta, ``Investigating the
  limits of randomized benchmarking protocols,'' \emph{Physical Review A},
  vol.~89, p. 062321, 2014.

\bibitem{asaad2016independent}
S.~Asaad, C.~Dickel, N.~K. Langford, S.~Poletto, A.~Bruno, M.~A. Rol,
  D.~Deurloo, and L.~DiCarlo, ``Independent, extensible control of
  same-frequency superconducting qubits by selective broadcasting,'' \emph{NPJ
  Quantum Information}, vol.~2, p. 16029, 2016.

\bibitem{heinsoo2018rapid}
J.~Heinsoo, C.~K. Andersen, A.~Remm, S.~Krinner, T.~Walter, Y.~Salath{\'e},
  S.~Gasperinetti, J.-C. Besse, A.~Poto{\v{c}}nik, C.~Eichler, and A.~Wallraff,
  ``Rapid high-fidelity multiplexed readout of superconducting qubits,''
  \emph{arXiv:1801.07904}, 2018.

\bibitem{reed2013entanglement}
M.~D. Reed, ``Entanglement and quantum error correction with superconducting
  qubits,'' Ph.D. dissertation, Yale University, 2013.

\bibitem{grover1996fast}
L.~K. Grover, ``A fast quantum mechanical algorithm for database search,'' in
  \emph{Proceedings of the 28th Annual ACM Symposium on Theory of
  Computing}.\hskip 1em plus 0.5em minus 0.4em\relax ACM, 1996, pp. 212--219.

\bibitem{dicarlo2009demonstration}
L.~DiCarlo, J.~M. Chow, J.~M. Gambetta, L.~S. Bishop, B.~R. Johnson, D.~I.
  Schuster, J.~Majer, A.~Blais, L.~Frunzio, S.~M. Girvin, and R.~J. Schoelkopf,
  ``Demonstration of two-qubit algorithms with a superconducting quantum
  processor,'' \emph{Nature}, vol. 460, pp. 240--244, 2009.

\bibitem{watson2018programmable}
T.~F. Watson, S.~G.~J. Philips, E.~Kawakami, D.~R. Ward, P.~Scarlino,
  M.~Veldhorst, D.~E. Savage, M.~G. Lagally, M.~Friesen, S.~N. Coppersmith,
  M.~A. Eriksson, and L.~M.~K. Vandersypen, ``A programmable two-qubit quantum
  processor in silicon,'' \emph{Nature}, vol. 555, p. 633, 2018.

\bibitem{cramer2016repeated}
J.~Cramer, N.~Kalb, M.~A. Rol, B.~Hensen, M.~S. Blok, M.~Markham, D.~J.
  Twitchen, R.~Hanson, and T.~H. Taminiau, ``Repeated quantum error correction
  on a continuously encoded qubit by real-time feedback,'' \emph{Nature
  Communications}, vol.~7, p. 11526, 2016.

\end{thebibliography}
\end{document}